\documentclass[11pt]{article}
\usepackage{graphicx}
\usepackage{amssymb,amsfonts,amsthm}
\usepackage{setspace}
\usepackage{booktabs,calc}
\usepackage{multirow}
\usepackage{float}
\usepackage{amsmath}
\usepackage[justification=centering]{caption}
\usepackage{subfig}
\usepackage{thmtools}
\usepackage{threeparttable}
\usepackage[title]{appendix}

\usepackage[textsize=tiny]{todonotes}
\usepackage{rotating}
\usepackage{enumitem}
\usepackage{listings}
\usepackage{color}
\usepackage{hyperref}

\makeatletter
\newcommand{\rmnum}[1]{\romannumeral #1}
\newcommand{\Rmnum}[1]{\expandafter\@slowromancap\romannumeral #1@}
\makeatother

\usepackage[top=1 in, bottom=1in, left=1 in , right=1in]{geometry}

\declaretheorem{theorem}
\declaretheorem{lemma}

\declaretheorem{assumption}
\declaretheorem[numbered=no,name=Assumption 1$'$]{assumption1'}

\declaretheorem[style=definition]{remark}

\newcounter{subassumption}[assumption]

\makeatletter
\renewcommand{\p@subassumption}{\theasu}
\makeatother

\numberwithin{equation}{section}
\numberwithin{theorem}{section}
\numberwithin{corollary}{section}
\numberwithin{assumption}{section}
\numberwithin{remark}{section}

\DeclareMathOperator*{\argmin}{arg\,min}

\def\lf{\lfloor}
\def\rf{\rfloor}

\makeatletter
\def\old@comma{,}
\catcode`\,=13
\def,{%
  \ifmmode%
    \old@comma\discretionary{}{}{}%
  \else%
    \old@comma%
  \fi%
}
\makeatother

\hypersetup{
    colorlinks,
    citecolor=black,%
    filecolor=black,%
    linkcolor=black,%
    urlcolor=black}

\begin{document}
\vspace{-15em}

\title{Inference in Nonparametric Series Estimation with Specification Searches for the Number of Series Terms}
\author{Byunghoon Kang\thanks{I thank the editor Peter Phillips, the co-editor Iv\'{a}n Fern\'{a}ndez-Val, and the two anonymous referees for thoughtful comments that significantly improved this paper. I am also grateful to Bruce Hansen, Jack Porter, Xiaoxia Shi and Joachim Freyberger for useful comments and discussions, and  thanks to Michal Koles\'{a}r, Denis Chetverikov, Yixiao Sun, Andres Santos, Patrik Guggenberger, Federico Bugni, Joris Pinkse, Liangjun Su, Myung Hwan Seo, and \'{A}ureo de Paula for helpful conversations and criticism. This paper is a revised version of the first chapter in my Ph.D. thesis at UW-Madison and previously titled ``Inference in Nonparametric Series Estimation with Data-Dependent Undersmoothing". I acknowledge support by the Kwanjeong Educational Foundation Graduate Research Fellowship and Leon Mears Dissertation Fellowship from UW-Madison. All errors are my own. \hangindent=1.5em Email: \href{mailto:b.kang1@lancaster.ac.uk}{b.kang1@lancaster.ac.uk}, Homepage: \href{https://sites.google.com/site/davidbhkang}{https://sites.google.com/site/davidbhkang}}\\
\normalsize{Department of Economics, Lancaster University} }

\vspace{-2em}
\date{\normalsize{\today}} 

\maketitle

\vspace{-2em}

\begin{abstract}
Nonparametric series regression often involves specification search over the tuning parameter, i.e.,  evaluating estimates and confidence intervals with a different number of series terms. This paper develops pointwise and uniform inferences for conditional mean functions in nonparametric series estimations that are uniform in the number of series terms. As a result, this paper constructs confidence intervals and confidence bands with possibly data-dependent series terms that have valid asymptotic coverage probabilities. This paper also considers a partially linear model setup and  develops inference methods for the parametric part uniform in the number of series terms. The finite sample performance of the proposed methods is investigated in various simulation setups as well as in an illustrative example, i.e., the nonparametric estimation of the wage elasticity of the expected labor supply from Blomquist and Newey (2002).
\newline

\textit{Keywords: } Nonparametric series regression, Pointwise confidence interval, Smoothing parameter choice, Specification search, Undersmoothing, Uniform confidence bands.

\textit{JEL classification: } C12, C14.
\end{abstract}

\section{Introduction} \label{sec:intro}

We consider the following nonparametric regression model
\begin{equation}\label{eq:model}
 y_i = g_0(x_i) + \varepsilon_i,\qquad  E(\varepsilon_i | x_i ) =   0
\end{equation}
where $\{ y_i, x_i  \}_{ i =1}^n$ is i.i.d.,  $y_i$ is a  scalar response variable, $x_i \in \mathcal{X} \subset  \mathbb{R}^{d_x}$ is a  vector of covariates, and $g_0(x) = E(y_i | x_i = x)$ is the conditional mean function. The theory of estimation and inference is well developed for nonparametric series (sieve) methods in a large body of  econometrics and statistics literature. Series estimators have also received attention in applied economics because they have many appealing features, e.g., they can easily impose shape restrictions  such as additive separability and  monotonicity. Once the basis function is chosen (e.g., polynomial or regression spline series of fixed order), implementation requires a choice of the number of series terms $K  = K_n$ where $K$ denotes the order of the polynomials or the number of knots in the splines. However, this often involves some ad-hoc specification searches over $K \in \mathcal{K}_n$. For example, when $x_i \in \mathbb{R}^{d_x}$ is vector valued, researchers often evaluate the different numbers of terms in each dimension separately and construct a set of bases with different powers and cross-products of covariates. Although specification search seems necessary in some cases, it may lead to misleading inference without considering the first-step specification search or series term selection.\footnote{As a referee noted, the bias and MSE of the series estimator depend on not only $K$ but also the specific bases or sieve spaces, e.g., the order of the splines. In this paper, we fix the basis function, and we do not allow searching over the specific bases or sieve spaces.}

Existing theory for the asymptotic normality of \textit{t}-statistics and valid inference imposes a so-called undersmoothing (i.e., overfitting) condition that is a faster rate of $K$ than the mean-squared error (MSE) optimal convergence rates, and many papers in the literature typically suggest rule-of-thumb rules that give the desired level of undersmoothing. Among many others, Newey (2013) suggested increasing $K$ until the standard errors are large relative to small changes in objects of interest. Newey, Powell, and Vella (1999) suggested using more terms than that chosen by cross-validation. Horowitz and Lee (2012) suggested increasing $K$ until the integrated variance suddenly increases and then adding additional terms. 

In this paper, we formally justify these rule-of-thumb methods  or ``plug-in'' methods with undersmoothed $\widehat{K}$ for valid inference in nonparametric series regression. Specifically, we provide pointwise inference for $g_0(x)$  with possibly data-dependent (undersmoothed) $\widehat{K}  \in \mathcal{K}_n$, i.e., constructing $100(1-\alpha) \%$ confidence interval (CI), 

\begin{equation}\label{def:pmsci}
\liminf\limits_{n \rightarrow \infty} P( g_0(x) \in  [ \widehat{g}_n(\widehat{K},x) \pm \widehat{c}_{1-\alpha} (x) \sqrt{\widehat{V}_n(\widehat{K}, x)/n} ]) \geq  1-\alpha,
\end{equation}
with an estimator $\widehat{g}_n(K,x)$, variance $\widehat{V}_n(K, x)$ using $K$ series terms, and critical values $\widehat{c}_{1-\alpha}(x)$ from the supremum of the \textit{t}-statistics. For this result, we first develop a uniform distributional approximation theory of the absolute value of the supremum of the \textit{t}-statistics over different series terms to construct asymptotically valid confidence intervals,  which are uniform in $K \in \mathcal{K}_n$, 
\begin{equation}\label{def:uniformci}
P( g_0(x) \in  [ \widehat{g}_n(K,x) \pm \widehat{c}_{1-\alpha} (x) \sqrt{\widehat{V}_n(K, x)/n} ], \quad   K \in \mathcal{K}_n) = 1-\alpha + o(1). 
\end{equation}
The critical values $\widehat{c}_{1-\alpha} (x)$ can be easily implemented using simple simulation or  weighted bootstrap methods. 

Furthermore, this paper develops the construction of confidence bands for $g_0(x)$ with asymptotically  uniform (in $K \in \mathcal{K}_n$) coverage with critical values $\widehat{c}_{1-\alpha} $ chosen to satisfy
\begin{equation} \label{def:uniformbands}
P( g_0(x) \in  [ \widehat{g}_n(K,x) \pm \widehat{c}_{1-\alpha} \sqrt{\widehat{V}_n(K, x)/n} ], \quad   K \in \mathcal{K}_n , x \in \mathcal{X})  =  1-\alpha + o(1).
\end{equation}
Analogous to the pointwise inference in \eqref{def:pmsci}, we can show the validity of confidence bands with the data-dependent $\widehat{K}$. Even in pointwise inference, deriving a uniform asymptotic distribution theory for all sequences of \textit{t}-statistics over $K \in \mathcal{K}$ may not be possible unless $p = |\mathcal{K}_n|$ is finite. Allowing $p \rightarrow \infty$ as $n \rightarrow \infty$, results in this paper build on coupling inequalities for the supremum of the empirical process developed by Chernozhukov, Chetverikov, and Kato (2014a, 2016)  combined with anti-concentration inequality in Chernozhukov, Chetverikov, and Kato (2014b).

We also provide inference methods in a partially linear model  setup focusing on the common parametric part.  Unlike the nonparametric object of interest that has a slower convergence rate than $n^{1/2}$ (e.g., regression function or regression derivative), the \textit{t}-statistics for the parametric object of interest are asymptotically equivalent for all sequences of $K$ under standard rate conditions $K/n \rightarrow 0$ as $n \rightarrow \infty$. To account for the dependency of the \textit{t}-statistics with the different sequences of $K$ in this setup, we consider a faster rate of $K$ that grows as fast as the sample size $n$, as in Cattaneo, Jansson, and Newey (2018a, 2018b), and develop an asymptotic distribution of the \textit{t}-statistics over $K \in \mathcal{K}_n$. Then, we discuss methods to construct confidence intervals that are similar to the nonparametric regression setup and provide uniform (in $K \in \mathcal{K}_n$) coverage properties. 

We investigate finite sample coverage and length properties of the proposed CIs and uniform confidence bands in various simulation setups. As an illustrative example, we revisit nonparametric estimation of labor supply function using the entire individual piecewise-linear budget set as in Blomquist and Newey (2002). Imposing additive separability, which is derived by economic theory, Blomquist and Newey (2002) estimate the conditional mean of labor supply function using series estimation and report wage elasticity of the expected labor supply as well as other welfare measures with various specifications of the different number of series terms. 

Several important papers have investigated the asymptotic properties of series (and sieve) estimators, including papers by Andrews (1991a); Eastwood and Gallant (1991); Newey (1997); Chen and Shen (1998); Huang (2003); Chen (2007); Chen and Liao (2014); Chen, Liao, and Sun (2014);  Belloni, Chernozhukov, Chetverikov, and Kato (2015); and Chen and Christensen (2015), among many others.  This paper extends inference based on the \textit{t}-statistic under a single sequence of $K$ to the sequences of $K$ over a set $\mathcal{K}_n$ and focuses both on the pointwise and uniform inferences on $g_0(x)$, which is irregular (i.e., slower than a rate of $n^{1/2}$) and a linear functional, under an i.i.d. setup.

The supremum \textit{t}-statistics have been used as a correction for multiple-testing problems and to construct simultaneous confidence bands, and the importance of multiple-testing problems (data mining or data snooping) has long been noted in various other contexts (see Leamer (1983), White (2000), Romano and Wolf (2005), Hansen (2005)).  

There is also a growing literature on data-dependent series term selection and its impact on estimation and inference in econometrics and statistics.  Asymptotic optimality results of cross-validation have been developed, e.g., in papers by Li (1987), Andrews (1991b), and Hansen (2015). Horowitz (2014) develops data-driven methods for choosing the sieve dimension in the nonparametric instrumental variables (NPIV) estimation such that resulting NPIV estimators attain the optimal sup-norm or $L^2$ norm rates adaptive to the unknown smoothness of $g_0(x)$.  Although we do not pursue adaptive inference in this paper, there is also a large statistical literature on adaptive inference. For example, Gin\'e and Nickl (2010), Chernozhukov, Chetverikov, and Kato (2014b) construct adaptive confidence bands in the density estimation problem  (see Gin\'e and Nickl (2015, Section 8) for comprehensive lists of references). However, once data-driven choice is obtained for adaptive estimation (e.g., Lepski (1990)-type procedures), one still requires an undersmoothing condition for inference to eliminate asymptotic bias terms (see Theorem 1 of Gin\'e and Nickl (2010)), and this may result in similar specification search issues when choosing sufficiently ``large" $K$ in practice. 

We can, in principle, consider kernel-based estimation where several data-dependent bandwidth selections or explicit bias corrections have been proposed.\footnote{See H\"{a}rdle and Linton (1994), Li and Racine (2007) for references. See also Hall and Horowitz (2013), Calonico, Cattaneo, and Farrell  (2018), Schennach (2015) and references therein for various recent works on related bias issues and inference for kernel estimators.} However, there exist many examples estimating $g_0(x)$ using (global) series estimation and imposing shape constraints easily (such as additive separability to reduce dimensionality) that are also interested in both pointwise and uniform inference. Given the issues of specification search, our paper is closely related to a recent paper by Armstrong and Koles\'{a}r (2018) which considers a bandwidth snooping adjustment for kernel-based inference.

Unlike kernel-based methods, little is known about the statistical properties of data-dependent selection rules and explicit bias formulas for general series estimation. Zhou, Shen, and Wolfe (1998) and Huang (2003) are two of the few exceptions. A recent paper, Cattaneo, Farrell, and Feng (2019), develops novel explicit asymptotic bias/integrated mean squared error (IMSE) formulas and asymptotic theory of the bias-correction methods for general partitioning-based series estimators. The results in Cattaneo, Farrel, and Feng (2019) can be used as an alternative to the undersmoothing approach to avoid specification search issues.

The remainder of the paper is organized as follows. Section \ref{sec:main} introduces the basic nonparametric series regression setup and the candidate set $\mathcal{K}_n$. Section \ref{sec:pointwise} provides the pointwise inference, and Section \ref{sec:uniform} provides uniform inference in $x \in \mathcal{X}$. Section \ref{sec:PLM} extends our inference methods to the partially linear model setup. Section \ref{sec:simulation} summarizes Monte Carlo experiments in various setups, and Section \ref{sec:application} illustrates an empirical example as in Blomquist and Newey (2002). Then,  Section \ref{sec:conclusion} concludes the paper. Appendix  \ref{sec:proof} includes the main proofs, and Appendix \ref{sec:figtable} includes figures and tables. Additional supporting lemmas and simulation results are provided in the Online Supplementary Material available at Cambridge Journals Online (journals.cambridge.org/ect).

\subsection{Notation}\label{sec:notation}
 $||A ||$ denotes the spectral norm, which equals the largest singular value of a matrix $A$, and $\lambda_{min}(A), \lambda_{max}(A)$ denote the  minimum and  maximum eigenvalues of a symmetric matrix $A$, respectively. $o_p(\cdot)$ and $O_p(\cdot)$ denote the usual stochastic order symbols, $\overset{d}{\longrightarrow}$ denotes convergence in distribution, and $\Rightarrow$ denotes weak convergence. Let $a \wedge b = \min \{ a,b \}, a \vee b = \max \{ a, b \}$ and denote $\lf a \rf$ as the largest integer less than the real number $a$. For two sequences of positive real numbers $a_n$ and $b_n$, $a_n \lesssim b_n$ denotes $a_n \leq c b_n$ for all $n$ sufficiently large with some constant $c>0$ that is  independent of $n$. $ a_n \asymp b_n$ denotes $a_n \lesssim b_n$ and $b_n \lesssim a_n$. Furthermore, $a_n \lesssim_{P} b_n$ denotes $a_n = O_p(b_n)$. For a given random variable $ \{ X_i \} $ and $1 \leq p <\infty$, $L^p(X)$ is the space of all $L^p$-norm bounded functions with $|| f ||_{L^p} = [E || f(X_i)||^p]^{1/p}$, $\ell^{\infty}(X)$ denotes the space of all bounded functions under the sup-norm, and $|| f ||_{\infty} = \sup_{x \in \mathcal{X}} |f(x)|$ for the bounded real-valued functions $f$ on the support $\mathcal{X}$.
 
\section{Setup} \label{sec:main}

We introduce the nonparametric series regression setup in the model \eqref{eq:model}. Given a random sample $\{ y_i, x_i  \}_{ i =1}^n$,  we are interested in inference on the conditional mean $g_0(x) = E(y_i | x_i = x)$ at a particular point $x \in \mathcal{X} \subset \mathbb{R}^{d_x}$ or uniform in $x \in \mathcal{X}$.

Let $\widehat{g}_{n} (K, x) $ be an estimator of $g_0(x)$ using $K =  K_n \geq 1$ series terms $P(K, x) = (p_{1}(x), \cdots, p_{K}(x) )'$, which is a vector of basis functions that can change with $n$. Standard examples for the basis functions are power series, Fourier series, orthogonal polynomials, splines and wavelets. 
The series estimator is then obtained by the least square (LS) estimation of $y_i$ on regressors $P (K, x_i)$
\begin{equation} \label{eq:estimator}
\widehat{g}_{n}(K, x) = P(K, x)' \widehat{\beta}_K, \qquad \widehat{\beta}_K  = (P^{K'} P^{K})^{-1} P^{K'} Y
\end{equation}
where  $P^K = [P_{K1}, \cdots , P_{Kn}]', P_{Ki} \equiv P(K, x_i) = (p_{1}(x_i), p_{2}(x_i), \cdots, p_{K}(x_i) )', Y = (y_1, \cdots y_n)' $. Define the least square residuals as $\widehat{\varepsilon}_{K i} = y_i - P_{K i}'\widehat{\beta}_K$, 
\begin{align} \label{def:varianceestimate}
\begin{split}
&\widehat{V}_n(K, x) = P(K, x)' \widehat{Q}_K^{-1} \widehat{\Omega}_K \widehat{Q}_K^{-1}  P(K, x),\\
& \widehat{Q}_K = \frac{1}{n}\sum_{i=1}^{n} P_{K i} P_{K i}', \quad  \widehat{\Omega}_K = \frac{1}{n}\sum_{i=1}^{n} P_{K i} P_{K i}' \widehat{\varepsilon}_{K i}^2, 
\end{split}
\end{align}
and consider the \textit{t}-statistic
\begin{equation} \label{eq:tstatistic}
\widehat{T}_n(K, x) \equiv \frac{\sqrt{n}(\widehat{g}_{n} (K, x) -g_0(x) )}{ \widehat{V}_{n} (K, x)^{1/2}}. 
\end{equation}

Under standard regularity conditions (discussed in the next section), the \textit{t}-statistic can be decomposed as follows: 
\begin{equation}\label{eq:t-decompose}
\widehat{T}_n(K,  x) =  \frac{1}{\sqrt{n}} \sum_{i=1}^{n} \frac{P (K, x)' Q_K^{-1} P_{ K i } \varepsilon_i}{\widehat{V}_{n}(K, x)^{1/2}}  -\frac{r_{n} (K, x)}{\sqrt{\widehat{V}_{n}(K, x)/n}}+  o_p(1)
\end{equation}
where $Q_K = E(P_{Ki} P_{Ki}')$, $r_n(K, x)= g_0(x) - P(K, x)' \beta_K$, and $\beta_K \equiv E[P_{Ki}P_{Ki}'])^{-1} E[P_{Ki}y_i]$ is the best linear $L^2$ projection coefficient. The first term in the decomposition \eqref{eq:t-decompose} converges to a standard normal distribution for the deterministic sequence $K \rightarrow \infty$ as $n\rightarrow \infty$, and the second term  does not necessarily converge to 0 due to approximation errors $ r_n(K, x)$. The second term can be ignored with an undersmoothing assumption, and the asymptotic distribution of the \textit{t}-statistic, $\widehat{T}_n(K, x) \overset{d}{\longrightarrow} N(0,1)$,  is well known in the literature (see, for examples, Andrews (1991a), Newey (1997), Belloni et al. (2015), and Chen and Christensen (2015), among many others). Then, the $100(1-\alpha) \%$ confidence interval for $g_0(x)$ can be easily constructed using the normal critical value $z_{1-\alpha/2}$
\begin{equation}\label{eq:standardCI}
\Big[ \widehat{g}_n(K,x) \pm z_{1-\alpha/2} \sqrt{\widehat{V}_n(K, x)/n} \Big].
\end{equation}

However, it is not clear whether the conventional CI using normal critical values \eqref{eq:standardCI} has a correct coverage probability with a possibly data-dependent $\widehat{K}$ such as cross-validation or IMSE-optimal selection. First, $T_n({\widehat{K}},\theta_0) \overset{d}{\rightarrow} N(0,1) $ may not hold with a random sequence of $\widehat{K}$, even if we assume the asymptotic bias is negligible. Second, it is well known that some data-dependent rules $\widehat{K}$ do not satisfy the undersmoothing rate conditions, which can lead to a large asymptotic bias and coverage distortion of the standard CI. For example, suppose that the researcher uses $\widehat{K} = \widehat{K}_{\texttt{cv}}$ selected by cross-validation; then, $\widehat{K}_{\texttt{cv}}$ is typically too ``small" and violates the undersmoothing assumption needed to ensure the asymptotic normality without bias terms and the valid inference.

As  discussed in the introduction, the undersmoothing assumption involves possibly ad-hoc methods to choose series terms $K$ over a \textit{candidate set} $\mathcal{K}_n$ for a valid inference, and cross-validation methods naturally involve specification search over a set of the different number of series terms.

The following  set assumption on $\mathcal{K}_n$ is constructed to allow a broad range of $K$ such that  $\mathcal{K}_n$ can allow (unknown) an optimal MSE rate of $K$ as well as an undersmoothing rate that increases faster than the optimal MSE rate.

\begin{assumption}\label{assump:sets_different} (Set of number of series terms) Assume the candidate set as $\mathcal{K}_n =  \{ K_j : 1\leq j \leq p \}$, where $ \underline{K} = K_1 \rightarrow \infty$ and $ \overline{K} = K_p \rightarrow \infty$ as $n\rightarrow \infty$. 
\end{assumption}

Here,  we consider a possibly growing set of the number of series terms, and a similar assumption is  used in the literature, for example, in  Newey (1994a, 1994b). Suppose $g_0(x)$ belongs to the H\"older space of smoothness $s >0$, $\Sigma (s, \mathcal{X})$; then, we obtain optimal $L^2$ convergence rates $O_p(n^{-s/(2s + d_x)})$ with $K \asymp n^{d_x/(d_x  + 2s)} $. Assumption \ref{assump:sets_different} allows having optimal $L^2$ rates of $K$ in a large set of classes of functions. By setting $\mathcal{K}_n = [\underline{K}, \overline{K}] \cap \mathbb{N}$,  $\overline{K} \asymp n^{ \overline{\phi}}$ and  $\underline{K} \asymp n^{ \underline{\phi}}$ with  $\overline{\phi} = d_x/(d_x  + 2\underline{s}), \underline{\phi} = d_x/(d_x  + 2\overline{s})$, Assumption \ref{assump:sets_different} contains the number of series terms that obtain an optimal $L^2$ rate of convergence for  $g_0(x) \in \bigcup_{s \in S} \Sigma (s, \mathcal{X}) $, $S = [\underline{s}, \overline{s}]$. A similar assumption is used in the literature on adaptive inference, although we do not pursue this direction in the current paper. 

Assumption \ref{assump:sets_different} gives flexible choices of $K$, as we only assume the rates of $K$, for example, $\overline{K} = C n^{\overline{\phi}}, \underline{K} = cn^{\underline{\phi}}$, where $c$ and $C$ can be set arbitrarily small or large. We only require rate restrictions uniformly over $K \in \mathcal{K}$ to guarantee the linearization of  the \textit{t}-statistic in \eqref{eq:t-decompose} and the rates of the cardinality $p = |\mathcal{K}_{n}|$. Since $K \in \mathcal{K}_n$ is a positive integer and $p \leq \overline{K}$, $p$ is growing at a rate much slower than $n$ under the rate restrictions in Section \ref{sec:pointwise}.

\begin{remark}[$\mathcal{K}_n$ and the largest $K$] 
As a referee noted, specification search is often performed over a simple pre-defined set in practice. For example, a researcher may only use quadratic, cubic, or quartic terms in polynomial regression or try only a few different numbers of knots in regression splines to observe how the estimate and standard error change. In the nonparametric estimation of the Mincer equation (Heckman, Lochner, and Todd (2006)), researchers may consider a regression of log wages on experience with polynomials of order $\underline{K}=1$ (linear) to  $\overline{K} = 4$ (quartic).\footnote{All of our results continue to hold with fixed $p$; however, it may be preferred to use larger sets $\mathcal{K}_n$ with $p \rightarrow \infty$ to give greater flexibility to the candidate models as the sample size $n$ increases.}

However, it may not be clear how to define \textit{a priori} $\mathcal{K}_n$ in practice.  One must first consider a set of pre-selected models over which to search. As discussed earlier and suggested by many papers in the literature, some formal data-dependent methods to obtain optimal  $L^2$ norm or sup-norm rates, such as cross-validation, can be a useful guideline for $\mathcal{K}_n$. For example, one can consider a reasonable set $\widetilde{\mathcal{K}}_n$ first, choose $ \widehat{K}_{\texttt{cv}} \in \widetilde{\mathcal{K}}_n$ by cross-validation, and then consider $\mathcal{K}_n = [\widehat{K}_{\texttt{cv}}, c_1  \widehat{K}_{\texttt{cv}} ]$ or $[ \widehat{K}_{\texttt{cv}},  \widehat{K}_{\texttt{cv}} n^{c_2}]$ for some constants $c_1, c_2>0$. One can also search $\underline{K}$ and $\overline{K}$ sequentially  by calculating changes in cross-validation or standard errors from the initial candidate set. Extending results developed in this paper with data-dependent $\mathcal{K}_n$ are beyond the scope of the paper.
\end{remark}

\section{Pointwise Inference}\label{sec:pointwise}

In this section, we focus on pointwise inference for $g_0(x)$. The goal of this section is to provide a uniform distributional approximation theory of $\widehat{T}_n(K,  x)$ over a set $\mathcal{K}_n$ and provide uniform (in $K \in \mathcal{K}_{n}$) coverage properties of confidence intervals for $g_0(x)$ in  \eqref{def:pmsci}, \eqref{def:uniformci} with the construction of critical values.

From the decomposition of the \textit{t}-statistic in \eqref{eq:t-decompose}, we first consider the (infeasible) test statistic 
\begin{equation}\label{def:infeasible-t}
\max_{K \in \mathcal{K}_{n}} |t_n(K,  x)| = \max_{ 1 \leq j \leq p} |t_n(K_j,  x)|
\end{equation}
where $t_n(K, x) =n^{-1/2} \sum_{i=1}^{n} P (K, x)' Q_K^{-1} P_{ K i } \varepsilon_i/ V_{n}(K, x)^{1/2}$ with the series variance $V_{n}(K, x) = P (K, x) ' Q_K^{-1} \Omega_K Q_K^{-1} P(K, x)$, $\Omega_K = E(P_{Ki} P_{Ki}' \varepsilon_{i}^2)$.  In general, $t_n(K, x), K\in \mathcal{K}_{n}$ does not have a limiting distribution because it is not asymptotically tight under Assumption \ref{assump:sets_different} unless $|\mathcal{K}_n|$ is finite or under the restrictive assumption on $\mathcal{K}_n$.\footnote{In an earlier version of the paper, we provide the weak convergence of a series process under the same rates of $K \in \mathcal{K}_n$ and high-level assumptions. This can be viewed as an analogous result in the kernel estimation literature (see Section 2 of Armstrong and Koles\'{a}r (2018) and other references therein).} However, we show below that there exists a sequence of random variables $\max_{ 1 \leq j \leq p} \sum_{i=1}^{n} |Z_{ij}|$ such that $\big|\max_{K \in \mathcal{K}_{n}} |t_n(K,  x)| - \max_{ 1 \leq j \leq p} \sum_{i=1}^{n} |Z_{ij}|\big| = O_p(a_n)$ for a sequence of constants $a_n \rightarrow 0$,  where $Z_{i} = (Z_{i1}, ..., Z_{ip})^{\prime}$ is a Gaussian random vector in $\mathbb{R}^{p}$ such that $Z_{i} \sim N(0, \frac{1}{n}\Sigma_{n})$ with $(j, l)$ elements of the variance-covariance matrix
\begin{equation}\label{def:covariance}
\Sigma_n (j, l) = E[t_n(K_j, x) t_n(K_l, x))] =  \frac{P(K_{j}, x)'  Q_{K_j}^{-1} \Omega_{K_j, K_l} Q_{K_l}^{-1} P(K_{l}, x) }{V_{n}(K_{j},x)^{1/2} V_{n} (K_{l},x)^{1/2}},
\end{equation}
$\Omega_{K_j, K_l} = E ( P_{ K_j i } P_{K_l i}' \varepsilon_{i}^2 )$.  

By replacing unknown  $\Sigma_n, V_n(K,x)$ with consistent estimators $\widehat{\Sigma}_{n}, \widehat{V}_{n}(K,x)$, we show below that we can approximate $\max_{K \in \mathcal{K}_{n}} |\widehat{T}_n(K,  x) |$ by $\max_{ 1 \leq j \leq p} \sum_{i=1}^{n} |Z_{ij}|$ and then
obtain critical values by using a simulation-based method to provide valid coverage properties in  \eqref{def:pmsci} and \eqref{def:uniformci}. We define $\widehat{c}_{1-\alpha} (x)$  as follows:
\begin{align}\label{def:cv-mcsup}
\begin{split}
&\widehat{c}_{1-\alpha} (x)  \equiv   (1-\alpha) \textnormal{  quantile of  } \max_{ 1 \leq j \leq p} \sum_{i=1}^{n} |\widehat{Z}_{ij}|, \textnormal{  where } \widehat{Z}_{i} = (\widehat{Z}_{i1}, ..., \widehat{Z}_{ip})^{\prime} \sim N(0, \frac{1}{n}\widehat{\Sigma}_{n}),\\
&  \widehat{\Sigma}_n (j, j) = 1, \quad \widehat{\Sigma}_n (j, l)=\frac{\widehat{V}_n(K_j, K_l,x) }{\widehat{V}_n (K_{j},x)^{1/2} \widehat{V}_n (K_{l},x)^{1/2}},\\
& \widehat{V}_n(K_j, K_l,x) = P(K_j, x)' \widehat{Q}_{K_j}^{-1} \widehat{\Omega}_{K_j, K_l} \widehat{Q}_{K_l}^{-1}  P(K_l, x), \widehat{\Omega}_{K_j, K_l} =  \frac{1}{n}\sum_{i=1}^{n} P_{K_j i} P_{K_l i}' \widehat{\varepsilon}_{K_j i} \widehat{\varepsilon}_{K_l i}
\end{split}
\end{align}
where  $\widehat{\Sigma}_n$ is a consistent estimator of the variance-covariance matrix $\Sigma_n$ defined in \eqref{def:covariance}, $\widehat{V}_n(K,x)$ is the simple plug-in estimator for $V_n(K,x)$ as in \eqref{def:varianceestimate}, and $\widehat{\varepsilon}_{K i} = y_i - P_{K i}'\widehat{\beta}_K, \forall K \in \mathcal{K}_n$. One can compute $\widehat{c}_{1-\alpha} (x)$ by simulating $B$ (typically $B=1000$ or $5000$) i.i.d. random vectors $\widehat{Z}_{i}^b \sim N(0,\frac{1}{n} \widehat{\Sigma}_n)$ and by taking a $(1-\alpha)$ sample quantile of $\{    \max \limits_{1 \leq j \leq p} \sum_{i=1}^{n} |\widehat{Z}_{ij}^{b}|: b=1, \cdots, B \}$. Alternatively, we can use weighted bootstrap methods. See Section \ref{sec:uniform} for the implementation and the validity of bootstrap procedures in the construction of confidence bands. 
 
To establish our main results, we impose mild regularity conditions uniform in $K \in \mathcal{K}_n$. For each $K \in \mathcal{K}_n$, define $\zeta_{K} \equiv \sup_{x \in \mathcal{X}} || P(K, x)||$ as the largest normalized length of the regressor vector and $\lambda_{K} \equiv (\lambda_{min} (Q_{K}))^{-1/2}$ for $K \times K$ design matrix $Q_K = E(P_{K i} P_{K i}')$.

\begin{assumption}\label{assump:model} (Regularity conditions - model)
\begin{enumerate}
\item[(\rmnum{1})] $\{y_i, x_i\}_{i=1}^{n}$ are $i.i.d$ random variables satisfying the model \eqref{eq:model}.
\item[(\rmnum{2})]  $\max \limits_{K \in \mathcal{K}_n}\lambda_{K} \lesssim 1$, and for each $K \in \mathcal{K}_n$, as $K \rightarrow \infty$, there exists $c_{K}, \ell_{K}$ such that
\begin{align*}
\sup_{x \in \mathcal{X}} |r_n(K,x) | \leq \ell_{K} c_{K},  \quad  E[r_n(K,x)^{2} ]^{1/2} \leq c_{K},
\end{align*}
where $r_n (K,x) = g_0(x) - P(K, x)'\beta_K, \beta_K = (E[P_{Ki}P_{Ki}'])^{-1} E[P_{Ki}y_i]$. 
\end{enumerate}
\end{assumption}

\begin{assumption}\label{assump:regularity} (Regularity conditions - pointwise inference)
\begin{enumerate}
\item[(\rmnum{1})]   $\max \limits_{K \in \mathcal{K}_n}  \sqrt{ \zeta_{K}^{2} \log K \log^{2} p  /n}(1+\sqrt{K}\ell_{K} c_{K}) + \ell_{K} c_{K} \log p \rightarrow 0$ as $n \rightarrow \infty$.

\item[(\rmnum{2})]$ \sup_{x \in \mathcal{X}} E (|\varepsilon_i|^3 | x_i=x) < \infty$, $ \inf_{x \in \mathcal{X}} E (\varepsilon_i^2 | x_i=x) > 0$, and either of the following conditions hold: (a) $\sup_{x \in \mathcal{X}}E[ |\varepsilon_{i}|^{q}  | x_i = x] < \infty$ for $q \geq 4$ or (b) there exists a constant $C>0 $ such that $\sup_{x \in \mathcal{X}} E[\exp(|\varepsilon_i|/C) | X_i = x] \leq 2$. 

\item[(\rmnum{3})] $\max \limits_{K \in \mathcal{K}_n} |\frac{V_n (K,x)}{\widehat{V}_n (K,x)} -1|  = o_p(1/\log p)$, $\max \limits_{1 \leq j, l \leq p} |\widehat{\Sigma}_{n}(j,l)  - {\Sigma}_{n}(j,l)| =o_p(1/\log^{2} p)$.
\end{enumerate}
\end{assumption}
\vspace{0.1cm}

Assumptions \ref{assump:model}(\rmnum{2}) and \ref{assump:regularity}(\rmnum{1}) are similar to those imposed in Belloni et al. (2015) and Chen and Christensen (2015), and all the discussions made there also apply here except that we impose rate conditions of $K$ uniformly over $\mathcal{K}_n$. The rate conditions can be replaced by the specific bounds of $\zeta_K, c_K, \ell_K$  with various sieve bases. For example, when $\mathcal{X} = [0,1]^{d_x}$,  the probability density of $x_i$ is uniformly bounded above and bounded away from zero, and $g_0(x) \in \Sigma (s, \mathcal{X})$, i.e., the H\"older space of smoothness $s>0$, then $\lambda_{K} \lesssim 1$, $\zeta_{K} \lesssim \sqrt{K}$, $\ell_K  c_K \lesssim K^{-(s \wedge s_0)/d_x}$ for regression spline series of order $s_0$, and Assumption \ref{assump:regularity}(\rmnum{1}) is satisfied when $ \sqrt{\overline{K} (\log^{3} \overline{K}) /n}(1+ \overline{K}^{1/2} \underline{K}^{-(s \wedge s_0)/d_x}) + \underline{K}^{-(s \wedge s_0)/d_x} \log \overline{K} \rightarrow 0$. Other standard regularity conditions in the literature (e.g., Newey (1997) and Chen (2007)) can also be used here, and the rate condition can be improved with different pointwise linearization and approximation bounds in Huang (2003) for splines and Cattaneo et al. (2019) for partitioning-based estimators.

Assumption \ref{assump:regularity}(\rmnum{2}) imposes either the bounded polynomial moment conditions or sub-exponential moments of the regression errors. Assumption \ref{assump:regularity}(\rmnum{3}) imposes the consistency of variance estimator $\widehat{V}_n(K,x)$  uniformly in $K \in \mathcal{K}_n$, and this holds under mild regularity conditions (see Lemma 5.1 of Belloni et al. (2015) and Lemma 3.1-3.2 of Chen and Christensen (2015)).

\vspace{0.2cm}

\begin{theorem} \label{thm:asymp-cov}
Suppose that Assumptions \ref{assump:sets_different}, \ref{assump:model}, and \ref{assump:regularity} hold and  that either of the  following rate conditions hold depending on the case (a) or (b) in  Assumption \ref{assump:regularity}(\rmnum{2}): (a) $(\max_{K} \zeta_K)^{2} \log^{5} n \log^3 p/n \vee \max_{K} \zeta_K \log^{3/4} n \log p/n^{1/2-1/q} \rightarrow 0$ or (b) $(\max_{K} \zeta_K)^{2} \log^{5} n \log^3 p/n \rightarrow 0$. If, in addition, we assume that $\max \limits_{K \in \mathcal{K}_n}|  \frac{\sqrt{n}r_n(K,x)}{V_{n}(K,x)^{1/2}} | = o(1/\sqrt{ \log p})$, then 
\begin{align}\label{result:Kolmogorov}
\sup_{u \in \mathbb{R}} \big| P(\max_{K \in \mathcal{K}_{n}} |\widehat{T}_n(K,  x) | \leq u) - P(\max_{ 1 \leq j \leq p} \sum_{i=1}^{n} |\widehat{Z}_{ij}| \leq u)  \big| = o(1), 
\end{align}
and the following coverage property holds  
\begin{align}\label{result:coverage-uniformci} 
P( g_0(x) \in  [ \widehat{g}_n(K,x) \pm \widehat{c}_{1-\alpha} (x) \sqrt{\widehat{V}_n(K, x)/n} ], \quad   K \in \mathcal{K}_n)  =  1-\alpha + o(1)
\end{align}
with a critical value $\widehat{c}_{1-\alpha}(x)$ defined in \eqref{def:cv-mcsup}. Alternatively, if we assume $|  \frac{\sqrt{n}r_n(\widehat{K},x)}{V_{n}(\widehat{K},x)^{1/2}} | = o(1/\sqrt{ \log p})$ with $\widehat{K} \in \mathcal{K}_n$, then the following holds:
\begin{equation}\label{result:coverage-pmsci} 
\liminf\limits_{n \rightarrow \infty} P( g_0(x) \in  [ \widehat{g}_n(\widehat{K},x) \pm \widehat{c}_{1-\alpha} (x) \sqrt{\widehat{V}_n(\widehat{K}, x)/n} ]) \geq  1-\alpha.
\end{equation}
\end{theorem}
\vspace{0.1cm}

Theorem \ref{thm:asymp-cov} provides a uniform coverage property of the confidence interval over $K \in \mathcal{K}_n$ for the regression function $g_0(x)$. Equation \eqref{result:coverage-pmsci} guarantees the asymptotic coverage of CI for data-dependent $\widehat{K} \in \mathcal{K}_n$ with undersmoothing. Note that standard inference methods in  the nonparametric regression setup typically consider a singleton set $\mathcal{K}_n = \{ K \} $ with $K \rightarrow \infty$ as $n \rightarrow \infty$. The rate restriction is mild because it only requires $\overline{K}/n^{1-2/q} \rightarrow 0$, up to $\log n$ terms, in case $(a)$ and $\overline{K}/n \rightarrow 0$, up to $\log n$ terms, in case $(b)$ when $\zeta_{K} \lesssim \sqrt{K}$ for splines and wavelet series. Theorem \ref{thm:asymp-cov} builds upon a coupling inequality for maxima of sums of random vectors in Chernozhukov, Chetverikov, and Kato (2014a) combined with the anti-concentration inequality in Chernozukhov, Chetverikov, and Kato (2014b). 

\begin{remark}[Undersmoothing assumption]\label{remark:withoutundersmooth}
Note that \eqref{result:coverage-uniformci} requires an undersmoothing assumption uniformly over $K \in \mathcal{K}_n$. Without $\max \limits_{K \in \mathcal{K}_n}|  \frac{\sqrt{n}r_n(K,x)}{V_{n}(K,x)^{1/2}} | = o(1)$, coverage in \eqref{result:coverage-uniformci} can be understood as the uniform confidence intervals for the pseudo-true value $g(K,x) = P(K, x)' \beta_K$, i.e., 
\begin{equation}\label{result:coverage-uniformci-pseudo}
P( g(K,x)  \in  [ \widehat{g}_n(K,x) \pm \widehat{c}_{1-\alpha} (x) \sqrt{\widehat{V}_n(K, x)/n} ], \quad   K \in \mathcal{K}_n)  =  1-\alpha + o(1)
\end{equation}
However, a uniform undersmoothing condition is not assumed in \eqref{result:coverage-pmsci}, and it only requires that the chosen $\widehat{K} \in \mathcal{K}_n$ satisfies the undersmoothing condition such that the asymptotic bias is negligible. This allows broader ranges of $K$ in $\mathcal{K}_n$ including an unknown optimal MSE rate. We formally justify rule-of-thumb methods for valid inference suggested in the literature that include an additional number of series terms,  a  blow up of the numbers after using cross-validation, or some ``plug-in" methods for choosing $\widehat{K}$ such as those in Newey, Powell, and Vella (1999),  Newey (2013). Here, uniform (in $K\in \mathcal{K}_n$) inference considers uncertainty from specification search and using larger critical values $\widehat{c}_{1-\alpha} (x)$ than the normal critical value $z_{1-\alpha/2}$.
\end{remark}

 \begin{remark}[Other functionals]
Here, we focus on the leading example with $g_0(x)$ for some fixed point $x \in \mathcal{X}$; however, we can consider other linear functionals $a(g_0(\cdot))$ such as the regression derivatives  $a(g_0(x)) = \frac{d}{dx} g_0(x)$. All the results in this paper can be applied to irregular (slower than $n^{1/2}$ rate) linear functionals using estimators  $ a(\widehat{g}_n(K, x)) = a_{K}(x)' \widehat{\beta}_K$ and an appropriate transformation of basis $a_K(x) = (a(p_1(x), \cdots, a(p_K(x)))' $ with proper smoothness condition on the functional and continuity conditions on the derivative as in Newey (1997). Although the verification of previous results for regular ($n^{1/2}$ rate) functionals, such as integrals and weighted average derivatives, is beyond the scope of this paper, we examine similar results for the partially linear model setup in Section \ref{sec:PLM}.
\end{remark}

\section{Uniform Inference}\label{sec:uniform}

This section provides construction of uniform confidence bands for  $g_0(x)$ (uniform in $K \in \mathcal{K}_{n}$) given in \eqref{def:uniformbands}. We define the following empirical process 
\begin{equation}
\widehat{T}_n(K, x) \equiv \frac{\sqrt{n}(\widehat{g}_{n} (K, x) -g_0(x) )}{\widehat{V}_{n} (K, x)^{1/2}} \end{equation}
over $\mathcal{K}_n \times \mathcal{X}$, and we show below that the supremum of the empirical process $\sup_{(K, x) \in \mathcal{K}_{n} \times \mathcal{X}} |\widehat{T}_n(K, x)|$ can be approximated by a sequence of random variables $ \sup_{(K, x) \in \mathcal{K}_{n} \times \mathcal{X}} |Z_n (K,x)|$, where $Z_n (K,x)$ is a tight Gaussian random process in $\ell^{\infty}( \mathcal{K}_{n} \times \mathcal{X})$ with zero mean and covariance function
\begin{equation}\label{def:covariance-uniform}
E[Z_n (K,x) Z_n (K^{\prime},x^{\prime})] =  \frac{P(K, x)'  Q_{K}^{-1} \Omega_{K, K^{\prime}} Q_{K^{\prime}}^{-1} P(K^{\prime}, x^{\prime}) }{V_{n}(K, x)^{1/2} V_{n} (K^{\prime},x^{\prime})^{1/2}}.
\end{equation}
Although the Gaussian approximation is an important first step, the covariance function \eqref{def:covariance-uniform} is generally difficult to construct for the purpose of uniform inference. Thus, we employ weighted bootstrap methods similar to Belloni et al. (2015) and show the validity of the bootstrap procedure for uniform confidence bands. 

Let $e_1, ..., e_n$ be a sequence of i.i.d. standard exponential random variables that are independent of $X^{n} = \{ x_1, ..., x_n \}$. For $(K,x) \in \mathcal{K}_n \times \mathcal{X}$, we define a (centered) weighted bootstrap process
\begin{equation}\label{def-bootstrap}
\widehat{T}_n^{e} (K,x) =\frac{\sqrt{n}(\widehat{g}_{n}^{e} (K, x) -\widehat{g}_{n} (K, x) )}{\widehat{V}_{n} (K, x)^{1/2}}
\end{equation}
where $\widehat{g}_{n}^{e}(K, x) = P(K, x)' \widehat{\beta}_{K}^{e}$, and $\widehat{\beta}_{K}^{e}$ is obtained by the following weighted least squares regression
\begin{equation}\label{def-weightedLS}
\widehat{\beta}_K^{e} = \argmin_{\beta \in \mathbb{R}^{K}} \sum_{i=1}^{n} e_i (y_i - P(K, x_i)^{\prime} \beta )^{2}.
\end{equation}
Define the critical value 
\begin{equation}\label{def:cv-mbuniform}
\widehat{c}_{1-\alpha} \equiv  (1-\alpha) \textnormal{ conditional quantile of  }  \sup_{K \in \mathcal{K}_n, x \in \mathcal{X}} |\widehat{T}_n^{e} (K,x) |  \textnormal{ given the data } X^{n}, 
\end{equation}
and we consider confidence bands of the form
\begin{equation}\label{def:uniformbands-sec4}
[ \widehat{g}_n(K,x) \pm \widehat{c}_{1-\alpha} \sqrt{\widehat{V}_n(K, x)/n} ], \quad   K \in \mathcal{K}_n, x \in \mathcal{X}.
\end{equation} 

 To provide the validity of the bootstrap critical values and confidence bands in \eqref{def:uniformbands-sec4}, we show below that the conditional distribution of $\sup_{(K, x) \in \mathcal{K}_{n} \times \mathcal{X}} |\widehat{T}_n^{e} (K,x)|$ is ``close" to the distribution of $ \sup_{(K, x) \in \mathcal{K}_{n} \times \mathcal{X}} |Z_n (K,x)|$ and that of $ \sup_{(K, x) \in \mathcal{K}_{n} \times \mathcal{X}} |\widehat{T}_n (K,x)|$ using coupling inequalities  for the supremum of the empirical process and the bootstrap process as in Chernozhukov et al. (2016). Then, similar to Theorem \ref{thm:asymp-cov}, this gives bounds on the Kolmogorov distance for the distribution functions of $P(\sup_{K \in \mathcal{K}_n, x \in \mathcal{X}} |\widehat{T}_n (K,x) | \leq u)$ and $P(\sup_{K \in \mathcal{K}_n, x \in \mathcal{X}} |\widehat{T}_n^{e} (K,x) | \leq u  | X^{n})$.
 
The following assumptions are used to establish the coverage probability of confidence bands uniformly over $K \in \mathcal{K}_{n}$. Define $\alpha(K,x) \equiv  Q_K^{-1/2} P(K,x) /V_n(K,x)^{1/2}$, and 
\[
\zeta^{L_1} = \max_{K \in \mathcal{K}_n} \sup_{x, x^{\prime}\in \mathcal{X}, x\neq x^{\prime}} \frac{||\alpha(K,x) - \alpha(K,x^{\prime})||}{||x- x^{\prime} ||}, \  \zeta^{L_2} =  \sup_{x\in \mathcal{X}} \max_{K, K^{\prime} \in \mathcal{K}_n: K \neq K^{\prime}}\frac{||\alpha(K,x) - \alpha(K^{\prime},x)||}{|K- K^{\prime} |}.
\]

\begin{assumption}\label{assump:regularity-uniform}(Regularity conditions - uniform inference)
\begin{enumerate}
\item[(\rmnum{1})]  $\sup_{x} E[ |\varepsilon_{i}|^{q}  | x_i = x] < \infty$ for $q \geq 4$ and  $ \inf_{x \in \mathcal{X}} E (\varepsilon_i^2 | x_i=x) > 0$.

\item[(\rmnum{2})] $\max_{K \in \mathcal{K}_n}\sqrt{\frac{\lambda_K^2 \zeta_K^2 \log K \log^{4} n}{n}} (n^{1/q}+ \ell_K c_K \sqrt{K}) + (\ell_{K} c_{K}) \log n \rightarrow 0$ as $n \rightarrow \infty$.

\item[(\rmnum{3})] $\log (\zeta^{L_1} \vee \zeta^{L_2}) \lesssim \log n$, $\max_{K} \zeta_K^{2q/(q-2)} \log^{3} n /n \lesssim 1$, and $\max_{K} \zeta_{K} \lesssim \log n$.

\item[(\rmnum{4})] $\sup \limits_{(K, x) \in \mathcal{K}_n \times \mathcal{X}} |\frac{V_n (K,x)}{\widehat{V}_n (K,x)} -1|  = o_p(1/\log^{2} n)$.
\end{enumerate}
\end{assumption}
\vspace{0.1cm}

For uniform inference, we require similar but slightly stronger conditions compared to Assumption \ref{assump:regularity}. We also impose mild rate restrictions on $\zeta^{L_1}, \zeta^{L_2}$ and $\max_{K\in\mathcal{K}_n} \zeta_K$ similar to Chernozhukov et al. (2014a) and Belloni et al. (2015). 

\begin{theorem} \label{thm:uniform} 
Suppose that Assumptions \ref{assump:sets_different}, \ref{assump:model}, and \ref{assump:regularity-uniform} hold, and $(\max_{K} \zeta_{K}) \log^{2+ 1/(2q)} n /n^{1/2-1/q} \rightarrow 0$, $(\max_K \zeta_K)^{2} \log^{7} n/n \rightarrow 0$ as $n\rightarrow \infty$.  If, in addition, we assume that $\sup \limits_{(K,x) \in \mathcal{K}_n\times \mathcal{X} }|  \frac{\sqrt{n}r_n(K,x)}{V_{n}(K,x)^{1/2}} | = o(1/\sqrt{ \log n})$, then
\begin{align}\label{result:coverage-uniformcb} 
P( g_0(x) \in  [ \widehat{g}_n(K,x) \pm \widehat{c}_{1-\alpha} \sqrt{\widehat{V}_n(K, x)/n} ], \quad   K \in \mathcal{K}_n , x \in \mathcal{X})  =  1-\alpha + o(1)
\end{align}
with a  critical value $\widehat{c}_{1-\alpha}$ in \eqref{def:cv-mbuniform}. 

Alternatively, if we assume $\sup \limits_{x \in \mathcal{X} }|  \frac{\sqrt{n}r_n(\widehat{K},x)}{V_{n}(\widehat{K},x)^{1/2}} | = o(1/\sqrt{ \log n})$ with $\widehat{K} \in \mathcal{K}_{n}$, then the following coverage property holds:
\begin{equation}\label{result:coverage-pmscb} 
\liminf\limits_{n \rightarrow \infty} P( g_0(x) \in  [ \widehat{g}_n(\widehat{K},x) \pm \widehat{c}_{1-\alpha} \sqrt{\widehat{V}_n(\widehat{K}, x)/n} ], \quad x \in \mathcal{X}) \geq  1-\alpha.
\end{equation}
\end{theorem}

\vspace{0.1cm}

Theorem \ref{thm:uniform} shows the uniform asymptotic coverage property of the confidence bands defined in  \eqref{def:uniformbands-sec4} uniformly over $K \in \mathcal{K}_n$. Furthermore, it shows a confidence band with possibly data-dependent $\widehat{K} \in \mathcal{K}_{n}$ having an asymptotic coverage of at least $1-\alpha$. The confidence band constructed in \eqref{result:coverage-pmscb}  requires a substantially weaker assumption on the undersmoothing similar to Theorem \ref{thm:asymp-cov}.

\section{Extension: Partially Linear Model}
\label{sec:PLM}
In this section we provide inference methods for the partially linear model (PLM) setup. 
For notational simplicity, we use similar notation as defined in the nonparametric regression setup. Suppose we observe random samples $\{ y_i, w_i , x_i  \}_{ i =1}^n$, where $y_i$ is the scalar response variable, $w_i \in \mathcal{W} \subset \mathbb{R}$ is the treatment/policy variable of interest, and  $x_i \in  \mathcal{X} \subset \mathbb{R}^{d_x}$ is a set of explanatory variables. For simplicity, we shall assume that $w_i$ is a scalar. We consider the model
\begin{align}\label{eq:model-PLM}
y_i = \theta_0 w_i + g_0(x_i) + \varepsilon_i, \qquad E(\varepsilon_i | w_i, x_i) =  0.
\end{align}
We are interested in inference on $\theta_0$ after approximating  an unknown function $g_0(x)$ by series terms/regressors $p(x_i)$ among a set of potential control variables. Specification searches can be performed  for the number of different approximating terms or for the number of covariates in estimating the nonparametric part.

The series estimator $\widehat{\theta}_{n}(K) $ for $\theta_0$ using the first $K = K_n$ terms is obtained by standard LS estimation of $y_i$ on $w_i$ and $P_{Ki} = P(K, x_i)$ and has the usual ``partialling out'' formula
\begin{align} \label{eq:estimator-PLM}
\widehat{\theta}_{n}(K)= \left( W'M_KW \right)^{-1} W'M_KY
\end{align}
where $W = (w_1, \cdots, w_n)', M_K =  I_K - P^K(P^{K'}P^K)^{-1}P^{K'}, P^K = [P_{K1}, \cdots, P_{Kn}]', Y = (y_1, \cdots, y_n)'$.  The asymptotic normality and valid inference for $\widehat{\theta}_{n}(K)$ have been developed in the literature.\footnote{See also Robinson (1988), Linton (1995) and references therein for the results of the kernel estimators.}  Donald and Newey (1994) derived the asymptotic normality of $\widehat{\theta}_{n}(K)$ under standard rate conditions $K/n \rightarrow 0$.   Belloni, Chernozukhov, and Hansen (2014) analyzed asymptotic normality and uniformly valid inference for the post-double-selection estimator even when $K$ is much larger than $n $ (see also Kozbur (2018)). Recent papers by Cattaneo, Jansson, and Newey (2018a, 2018b) provided a valid approximation theory for $\widehat{\theta}_{n}(K)$ when $K$ grows at the same rate of $n$.

A different approximation theory using a faster rate of $K$ ($K/n \rightarrow c, 0<c <1 $) than the standard rate conditions ($K/n \rightarrow 0$) is particularly useful for our purpose to establish the asymptotic distribution of \textit{t}-statistics over $K \in \mathcal{K}_n$. From the results in Cattaneo, Jansson, and Newey (2018a), we have the following decomposition:
\begin{align} 
\nonumber\sqrt{n} (\widehat{\theta}_n(K) - \theta_0) &= ( \frac{1}{n} W'M_K W)^{-1} \frac{1}{\sqrt{n}} W' M_K Y \\
&=  \widehat{\Gamma}_n(K)^{-1} ( \frac{1}{\sqrt{n}} \sum_{i}v_i M_{K, ii}  \varepsilon_i  + \frac{1}{\sqrt{n}} \sum_{i=1}^{n} \sum_{j=1, j \neq i}^n v_i M_{K, ij}  \varepsilon_j )+ o_p(1) \label{eq:t-decompose-PLM}
\end{align}

\noindent where $v_i  \equiv w_i - g_{w0}(x_i) $,  $g_{w0}(x_i ) \equiv E [w_i | x_i]$ and $ \widehat{\Gamma}_n(K) = W'M_K W/n$. For any deterministic sequence $K \rightarrow \infty$ satisfying standard rate conditions $K/n \rightarrow 0$, $\sqrt{n} (\widehat{\theta}_n(K) - \theta_0) $ is asymptotically normal with variance $V = \Gamma^{-1} \Omega \Gamma^{-1}, \Gamma = E[v_i v_i^{\prime}], \Omega = E[v_i v_i^{\prime} \varepsilon_{i}^{2}]$.  Unlike the nonparametric object of interest in the fully nonparametric model, where the variance term increases with $K$, $\widehat{\theta}_{n}(K)$ has a parametric ($n^{1/2}$) convergence rate, and $\widehat{\theta}_{n}(K)$ with all different sequences of $K$ are asymptotically equivalent under $K/n \rightarrow 0$.\footnote{This is also related to the well-known results of the two-step semiparametric estimation; the asymptotic variance of two-step semiparametric estimators does not depend on the type of the first-step estimator or smoothing parameter sequences under certain conditions (see Newey (1994b)).} However, under faster rate conditions, $K/n \rightarrow c$ for $0<  c <1 $, the second term in \eqref{eq:t-decompose-PLM} is not negligible and converges to bounded random variables. Cattaneo, Jansson, and Newey (2018a) apply the central limit theorem of degenerate U-statistics for the second term, similar to the many instrument asymptotics analyzed in Chao, Swanson, Hausman, Newey, and Woutersen (2012). Then, the limiting normal distribution has a larger variance than the standard first-order asymptotic variance, and the adjusted variances generally depend on the number of terms $K$ such that we can provide an asymptotic distribution of the \textit{t}-statistics with the different sequence of $K$ over $\mathcal{K}_n$.

The following assumption on $\mathcal{K}_n$ is considered,  and we impose the regularity conditions that are used in Cattaneo, Jansson, and Newey (2018a, Assumption PLM) uniformly over $K \in \mathcal{K}_n$. 

\begin{assumption}\label{assump:sets-finite} (Set of finite number of series terms)
\begin{enumerate}
\item[] Assume $\mathcal{K}_n = \{ \underline{K} \equiv K_1 , \cdots, K_m, \cdots, \overline{K} \equiv K_p \}$, where  $K_m  \rightarrow \infty, K_m/n \rightarrow c_m$ as $n \rightarrow \infty$ for all $m=1, ... , p$, constant $c_m$ such that $0< c_1 < c_2 < \cdots < c_p  < 1$, and fixed $p$.
\end{enumerate}
\end{assumption}

\begin{assumption} \label{assump:PLM} (Regularity conditions - partially linear model)
\begin{enumerate}
\item[(\rmnum{1})] $\{y_i,w_i, x_i\}_{i=1}^{n}$ are $i.i.d$ random variables satisfying the model \eqref{eq:model-PLM}.
\item[(\rmnum{2})] There exists constants $0<c \leq C<\infty$ such that $E[\varepsilon_i^2 |w_i, x_i] \geq c$ and $E[v_i^2 |x_i] \geq c$, $E[\varepsilon_i^4 |w_i, x_i] \leq C$ and $E [v_i^4 | x_i] \leq C$.
\item[(\rmnum{3})] $rank(P_{K}) = K$(a.s.) and $M_{K,ii} \geq C$ for $C >0$ for all $K \in \mathcal{K}_n$.
\item[(\rmnum{4})] For each $K \in \mathcal{K}_n$, there exists some $\gamma_g, \gamma_{g_w}$,
\begin{align*}
\min_{\eta_g} E[(g_0(x_i) - \eta_g' P_{K i})^2] = O(K^{-2\gamma_g}), \quad \min_{\eta_{g_w}} E[(g_{w0}(x_i) - \eta_{g_w}' P_{K i})^2] = O(K^{-2\gamma_{g_w}}).
\end{align*}
\end{enumerate}
\end{assumption}
\vspace{0.1cm}
Assumption \ref{assump:PLM} does not require $K/n \rightarrow 0$ which is required to obtain asymptotic normality in the literature (e.g., Donald and Newey (1994)). Similar to Assumption \ref{assump:regularity}(\rmnum{3}) in the nonparametric setup,  Assumption \ref{assump:PLM}(\rmnum{4}) holds for the polynomials and spline basis.  For example, \ref{assump:PLM}(\rmnum{4}) holds with $\gamma_g =s_g/d_x, \gamma_{g_w} = s_w/d_x $ when $\mathcal{X}$ is compact and when the unknown functions $g_0(x)$ and $g_{w0}(x)$ have $s_g$ and $s_w$ continuous derivates, respectively. 

Under Assumptions \ref{assump:sets-finite}, \ref{assump:PLM} and undersmoothing condition  ($n\overline{K}^{-2(\gamma_g+ \gamma_{g_w})} \rightarrow 0$), we have a joint asymptotic distribution of the \textit{t}-statistics $T_n (K, \theta) = \sqrt{n} V_{n}(K)^{-1/2}(\widehat{\theta}_n(K) - \theta_0)$  over $K \in \mathcal{K}_n$:
\begin{equation*}
(T_n(K_1, \theta_0), \cdots, T_n(K_p, \theta_0))' \overset{d}{\longrightarrow} Z_{\Sigma} = (Z_1, \cdots Z_p)' \sim N(0, \Sigma)
\end{equation*}
\noindent where 
\begin{align*}
 V_n(K) &= \Gamma_{n}(K)^{-1} \Omega_n(K) \Gamma_{n}(K)^{-1},\\
 \Gamma_n(K) &= \frac{1}{n} \sum_{i=1}^{n} M_{K, ii} E[v_i^2 | x_i],  \ \Omega_n(K) =  \frac{1}{n} \sum_{i=1}^{n}  \sum_{j=1}^{n} M_{K, ij}^{2} E[v_i^2 \varepsilon_{j}^{2}| x_i, x_j], 
\end{align*}
and the variance-covariance matrix $\Sigma$ with $(l, l^{\prime})$ element
\begin{align}
\label{def:covariance_plm}
\begin{split}
&\Sigma (l, l^{\prime}) \equiv \lim_{n \rightarrow \infty} \frac{V_n(K_{l}, K_{l^{\prime}})}{V_n (K_{l})^{1/2}(K_{l^{\prime}})^{1/2}}, \quad V_n(K_{l}, K_{l^{\prime}}) = \Gamma_{n}(K_{l})^{-1} \Omega_n(K_{l}, K_{l^{\prime}}) \Gamma_{n}(K_{l^{\prime}})^{-1}\\
&  \Omega_n(K_{l}, K_{l^{\prime}}) = \frac{1}{n} \sum_{i=1}^{n}  \sum_{j = 1}^{n} M_{K_{l}, ij}M_{K_{l^{\prime}}, ij} E[v_i^2 \varepsilon_{j}^{2}| x_i, x_j],
\end{split} 
\end{align}
for $l, l^{\prime} = 1, ..., p.$ Then, we can similarly define critical values as in \eqref{def:cv-mcsup} to construct confidence intervals for $\theta_0$ uniform in $K \in \mathcal{K}_{n}$ analogous to the nonparametric setup. Let  
\begin{equation}
\label{def:cv-mcsup_plm}
\widehat{c}_{1-\alpha} \equiv (1-\alpha)  \textnormal{  quantile of  } \max_{m=1, ..., p} |\widehat{Z}_{m}|,\ \ \widehat{Z}_{\Sigma} = (\widehat{Z}_{1}, ..., \widehat{Z}_{p})^{\prime} \sim N(0, \widehat{\Sigma}_{n})
\end{equation}
where $\widehat{\Sigma}_{n}$ is a consistent estimator for unknown $\Sigma$ defined in \eqref{def:covariance_plm}.

Theorem \ref{thm:PLM} is the main result for the partially linear model setup and provides the asymptotic coverage results of the CIs uniform in $K \in \mathcal{K}_n$ analogous to the nonparametric setup in Section \ref{sec:pointwise}. 

 \vspace{0.2cm}
\begin{theorem} \label{thm:PLM}
Suppose that Assumptions \ref{assump:sets-finite} and \ref{assump:PLM}  hold. In addition, assume that $n\overline{K}^{-2(\gamma_g+ \gamma_{g_w})} \rightarrow 0$ and $\max \limits_{K, K^{\prime} \in \mathcal{K}_n} |\frac{\widehat{V}_n (K, K^{\prime})}{V_n (K, K^{\prime})} -1|  = o_p(1)$ as $n, K \rightarrow \infty$. Then, 
\begin{align}\label{result:coverage-plmci} 
&\lim\limits_{n \rightarrow \infty} P( \theta_0 \in [ \widehat{\theta}_{n}(K) \pm \widehat{c}_{1-\alpha} \sqrt{\widehat{V}_{n}(K)/n}], \quad \forall  K \in \mathcal{K}_n )  =  1-\alpha,\\
\label{result:coverage-plmci-pms}&\liminf\limits_{n \rightarrow \infty} P( \theta_0 \in   [ \widehat{\theta}_{n}(\widehat{K}) \pm \widehat{c}_{1-\alpha} \sqrt{\widehat{V}_{n}(\widehat{K})/n}]) \geq  1-\alpha, \quad \widehat{K} \in \mathcal{K}_n,
\end{align}
where the critical value $\widehat{c}_{1-\alpha}$ is defined in \eqref{def:cv-mcsup_plm}. 
\end{theorem}
\vspace{0.1cm}

\begin{remark}
Note that the construction of CIs requires consistent variance estimation of $\Omega_n(K)$. As discussed in Cattaneo, Jansson, and Newey (2018a, 2018b), the construction of the heteroskedasticity-robust estimator for $\Omega_n(K)$ under $K/n \rightarrow c > 0$ is challenging, and the Eicker-Huber-White-type variance estimator generally requires $K/n \rightarrow 0$ for consistency. Cattaneo, Jansson, and Newey (2018b) considers the following standard error formula:
\begin{equation}
\widehat{\Omega}_n(K, \kappa_n) =  \frac{1}{n} \sum_{i=1}^{n}  \sum_{j=1}^{n} \kappa_{ij} \hat{v}_{K,i}^2 \hat{\varepsilon}_{K, j}^{2}
\end{equation}
where $\hat{v}_{K} = M_K W, \hat{\varepsilon}_{K} = M_{K}(Y - W\widehat{\theta}_n(K))$ and symmetric matrix $\kappa_n$ with $(i, j)$ element $\kappa_{ij}$. Cattaneo, Jansson, and Newey (2018b) show that $\widehat{\Omega}_n(K, \kappa_n)$ is consistent even under heteroskedasticity and $K/n \rightarrow c > 0$ with a certain choice of $\kappa_n$ and provide a  sufficient condition for consistency. See Theorems 3 and 4 of Cattaneo, Jansson, and Newey (2018b) for further discussion.
\end{remark}

\section{Simulations} \label{sec:simulation}
This section investigates the small sample performance of the proposed inference methods. We report the empirical coverage and the average length of the confidence intervals/confidence bands considered in Sections \ref{sec:pointwise} and \ref{sec:uniform} with various simulation setups. 

\begin{figure}[H]
    \centering
     \includegraphics[width=0.9\textwidth]{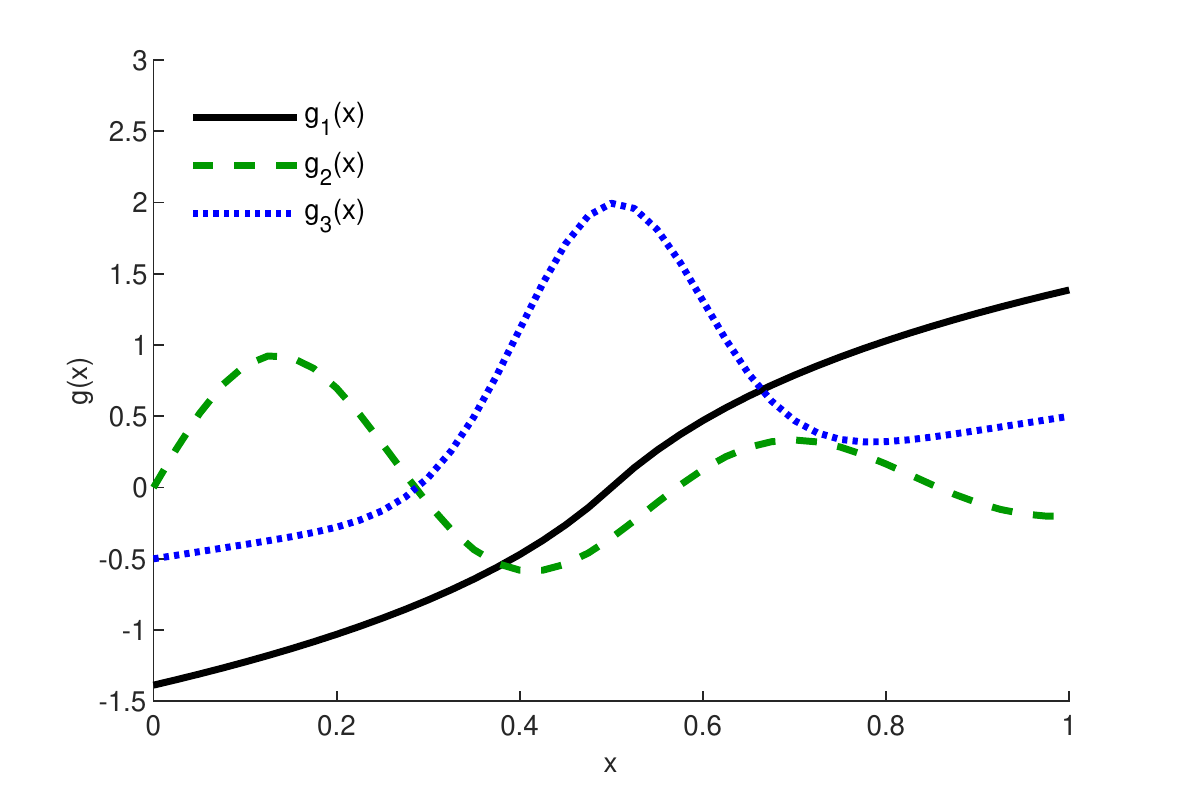}
        \caption{Different functions of $g(x)$ used in simulations (Section \ref{sec:simulation}). \\ Solid lines (Black) are $g_1(x) = \ln (|6x-3| + 1)sgn(x-1/2)$; Dashed lines (Green) are $g_2(x) = \sin (7\pi x/2)/ [1+ 2x^2(sgn(x)+1)]$ ; Dotted lines (Blue) are $g_3(x) = x-1/2 + 5 \phi(10(x-1/2))$, where $\phi(\cdot)$ is the standard normal pdf.}
    \label{fig:gfunctions}
\end{figure}

We consider the following data generating process:
\begin{align*}
y_i &= g(x_i) + \varepsilon_i, \\
x_i = \Phi(x_{i}^{*}), \begin{pmatrix}
x_i^* \\ \varepsilon_i
\end{pmatrix}
&\sim N
\begin{pmatrix}
\begin{pmatrix}
0 \\ 0
\end{pmatrix}, \begin{pmatrix}
1 & 0\\
0 & \sigma^2(x_{i}^{*})
\end{pmatrix}
\end{pmatrix}
\end{align*}
where $\Phi(\cdot)$ is the standard normal cumulative distribution function needed to ensure compact support, and $\sigma^2(x_{i}^{*})= ((1+ 2x_{i}^{*})/2)^{2}$ (heteroskedastic). We investigate the following three functions for $g(x)$:  $g_1(x) = \ln (|6x-3| + 1)sgn(x-1/2)$,  $g_2(x) = \frac{\sin (7\pi x/2)}{1+ 2x^2(sgn(x)+1)}$, and $g_3(x) = x-1/2 + 5 \phi(10(x-1/2))$, where $\phi(\cdot)$ is the standard normal probability density function, and $sgn(\cdot)$ is the sign function.  $g_1(x)$ is used in Newey and Powell (2003), as well as Chen and Christensen (2018). $g_2(x)$ and $g_3(x)$ are rescaled versions used in Hall and Horowitz (2013).  See Figure \ref {fig:gfunctions} for the shapes of all three functions on $[0,1]$. For all simulation results below, we generate 2000 simulation replications for each design with a sample size $n = 200$. 

 Results for quadratic splines with evenly placed knots are reported where the number of knots $K$ are selected among  $\mathcal{K}_n  = \{ 6, 7, ..., 12\}$ by setting $\underline{K} = 2n^{1/5}$ and $\overline{K} =  2n^{1/3}$ rounded up to the nearest integer. Then, we calculate a pointwise coverage rate (COV) and the average length (AL) of various 95\% nominal CIs, as well as analogous uniform CBs for the grid points of $x$ on the support $\mathcal{X} = [0.05,0.95]$. To calculate critical values, 1000 additional Monte Carlo or bootstrap replications are performed on each simulation iteration. In addition, we investigate results for homoskedastic errors ($\sigma^2(x_{i}^{*})= 1$), different sample sizes $n = \{ 100, 500 \}$, polynomial regressions, and different specifications as in Cattaneo and Farrell (2013) with multivariate and non-normal regressors; however, the results show qualitatively similar patterns  and hence are not reported here for brevity. Additional simulation results are reported in the Online Supplementary Material.

Table \ref{table:cov-spline-n1000} reports the nominal 95\% coverage of the following pointwise CIs at $x=0.2, 0.5, 0.8, 0.9$: (1) the standard CI in \eqref{eq:standardCI} with $\widehat{K}_{\texttt{cv}} \in \mathcal{K}_n$ selected to minimize the leave-one-out cross-validation; (2) robust CI in \eqref{result:coverage-pmsci}  with $\widehat{K}_{\texttt{cv}}$ using the critical value $\widehat{c}_{1-\alpha} (x)$; (3) robust CI using $\widehat{K}_{\texttt{cv+}} = \widehat{K}_{\texttt{cv}} + 2$. Analogous uniform inference results for CBs are also reported. The critical values,  $\widehat{c}_{1-\alpha} (x)$ and $\widehat{c}_{1-\alpha}$ are constructed using the Monte Carlo methods and weighted bootstrap method, respectively.

Overall, we find that the coverage of the standard CI with $\widehat{K}_{\texttt{cv}}$ is far less than 95\% over the support although it has the shortest length. However, the coverage of robust CIs based on  $\widehat{K}_{\texttt{cv}}$ or  $\widehat{K}_{\texttt{cv+}}$ with $\widehat{c}_{1-\alpha} (x)$  is close to or above 95\% and performs well across the different simulation designs, and this is consistent with theoretical results in Theorem \ref{thm:asymp-cov}. Using the undersmoothed $\widehat{K}_{\texttt{cv+}}$ (using more terms than the cross-validation) seems to work quite well at most points and for highly nonlinear designs where there exists relatively large bias, e.g., Model 3 $(g_{3}(x))$ at $x=0.5$.\footnote{The possibly poor coverage property of the standard kernel-based CIs for $g_3(x)$ at the single peak ($x= 0.5$) was also described in Hall and Horowitz (2013, Figure 3).} Uniform coverage rates of confidence bands with selected $K$ seem conservative, and this is due to the large critical values based on weighted bootstrap methods to be uniform in both $K \in \mathcal{K}_n$ and $x \in \mathcal{X}$, including boundary points.

\section{Empirical application} \label{sec:application}
In this section, we illustrate inference procedures by revisiting  Blomquist and Newey (2002). Understanding how tax policy affects individual labor supply has been a central issue in labor economics (see Hausman (1985) and Blundell and MaCurdy (1999), among many others). Blomquist and Newey (2002) estimate the conditional mean of hours of work given the individual nonlinear budget sets using nonparametric series estimation. They also estimate the wage elasticity of the expected labor supply and find  evidence of possible misspecification of the usual parametric model such as maximum likelihood estimation (MLE). 

Specifically, Blomquist and Newey (2002) consider the following model by exploiting an additive structure from the utility maximization with piecewise linear budget sets:
\begin{align}
\label{emp:model} h_i &= g(x_i) + \varepsilon_i, \quad E(\varepsilon_i | x_i) = 0,\\
\label{emp:mean}  g(x_i) &= g_1(y_J, w_J) + \sum_{j = 1}^{J-1} [g_2(y_j, w_j, \ell_j) - g_2(y_{j+1}, w_{j+1}, \ell_j)],
\end{align}
where $h_i$ is the hours worked of the $i$th individual and $x_i = (y_1, \cdots, y_J, w_1, \cdots, w_J, \ell_1, \cdots, \ell_J, J)$ is the budget set, which can be represented by the intercept $y_j$ (non-labor income), slope $w_j$ (marginal wage rates) and the end point $\ell_j$ of the $j$th segment in a piecewise linear budget with $J$ segments.  Equation \eqref{emp:mean} for the conditional mean function follows from Theorem 2.1 of Blomquist and Newey (2002), and this additive structure substantially reduces the dimensionality issues. To approximate $g(x)$, they consider the power series, $p_k(x) = (y_J^{p_1(k)} w_J^{q_1(k)}, \sum_{j=1}^{J-1} \ell_j^{m(k)} (y_j^{p_2(k)} w_j^{q_2(k)} - y_{j+1}^{p_2(k)} w_{j+1}^{q_2(k)})),$ $ p_2(k) + q_2(k) \geq 1$.

From the Swedish ``Level of Living" survey in 1973, 1980 and 1990, they pool the data from three waves and use the data for married or cohabiting men of ages 20-60. Changes in the tax system over three different time periods give a large variation in the budget sets. The sample size is $n = 2321$. See Section 5 of Blomquist and Newey (2002) for more detailed descriptions. They estimate the wage elasticity of the expected labor supply
\begin{equation}
E_w = \bar{w}/\bar{h} [\frac{\partial g(w, \cdots, w, \bar{y}, \cdots, \bar{y}) }{\partial w}] |_{w = \bar{w}},
\end{equation}
which is the regression derivative of $g(x)$ evaluated at the mean of the net wage rates $\bar{w}$, virtual income $\bar{y}$ and level of hours $\bar{h}$.

Table \ref{table:BN} is the same table as in Blomquist and Newey (2002, Table 1). They report estimates $\widehat{E}_w$ and standard errors $SE_{\widehat{E}_w}$ with a different number of series terms by adding additional series terms. For example, the estimates in the second row use the term in the first row $(1, y_J, w_J)$ with the additional terms $(\Delta y, \Delta w)$. Here, $\ell^m \Delta y^p w^q$ denotes approximating the term $\sum_j \ell_j^{m} (y_j^{p} w_j^{q} - y_{j+1}^{p} w_{j+1}^{q})$. Blomquist and Newey (2002) also report cross-validation criteria, $CV$, for each specification. In their formula, series terms are chosen to maximize $CV$, which minimizes the asymptotic MSE.  In addition to their original table, we add the standard 95\% CI for each specification, i.e., $CI (K) = \widehat{E}_w (K) \pm 1.96 SE_{\widehat{E}_w} (K)$. In Table \ref{table:BN}, it is ambiguous as to which large model ($K$) can be used for the inference, and we do not have compelling data-dependent methods for selecting one of the large $K$ for the confidence interval to be reported. Here we want to construct CIs that are robust to specification searches. 

\begin{figure}[H]
    \centering
    \caption{Nonparametric Wage Elasticity of Hours of Work Estimates in Blomquist and Newey (Table 1, 2002).}
     \includegraphics[width=0.85\textwidth]{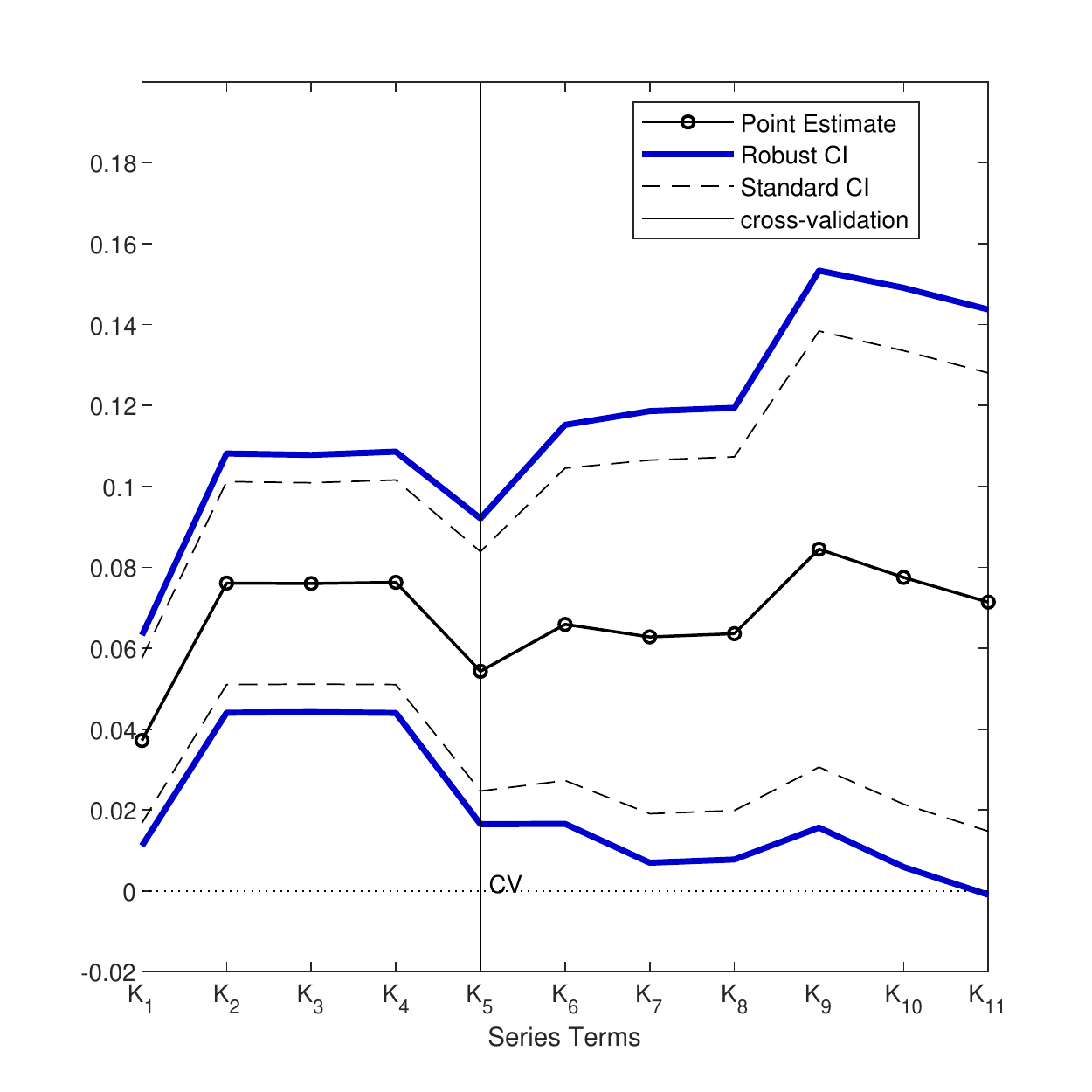}
        
    \label{fig:uniformCI}
    \vspace{0.4cm}
\begin{minipage}{0.85\textwidth} 
\centering
\small Figure \ref{fig:uniformCI} plots the wage elasticity  estimates  of the expected labor supply same as in Table \ref{table:BN}, with standard pointwise 95\% CIs as well as uniform (in $K \in \mathcal{K}_n$) CIs constructed with the critical value $\widehat{c}_{1-\alpha}(x)$.
\end{minipage}
\end{figure}

Figure \ref{fig:uniformCI} displays pointwise 95\% uniform CIs for $K_m \in \{ K_1, K_2, \cdots, K_{11}\}$, where $K_m$ corresponds to each specification in Table \ref{table:BN} with increasing order of series terms, along with the point estimates and standard 95\% confidence interval.\footnote{It is straightforward to construct $\widehat{c}_{1-\alpha}(x)$ using the covariance structure under the homoskedastic error and it only requires estimated variances for different $K\in \mathcal{K}_n$ that are already reported in the table of Blomquist and Newey (2002). Based on 100,000 simulation repetitions, we have $\widehat{c}_{1-\alpha}(x) = 2.503$.} From Figure \ref{fig:uniformCI}, we reject a zero wage elasticity of the labor supply for almost all models except $\overline{K}$. Table \ref{table:BN} also reports robust confidence intervals $CI_{\widehat{E}_w}^{\texttt{sup}} (K) = \widehat{E}_w (K) \pm \widehat{c}_{1-\alpha}(x) SE_{\widehat{E}_w} (K)$ with possibly data-dependent $\widehat{K}$ justified by Theorem \ref{thm:asymp-cov} (eq \eqref{result:coverage-pmsci}). Note that cross-validation chooses $ \widehat{K}_{\texttt{cv}} = K_5$, and the standard CI with $\widehat{K}_{\texttt{cv}}$ is $[0.0247, 0.0839]$ and the robust CI is $[0.0165, 0.0921]$. Using $\widehat{K}_{\texttt{cv+}} = K_6$ or $\widehat{K}_{\texttt{cv++}} = K_7$ widens the standard CI, and the robust CIs are $CI_{\widehat{E}_w}^{\texttt{sup}} (\widehat{K}_{\texttt{cv+}})= [0.0166, 0.1152], CI_{\widehat{E}_w}^{\texttt{sup}} (\widehat{K}_{\texttt{cv++}})  = [0.0070, 0.1186]$.

\section{Conclusion}\label{sec:conclusion}

This paper considers nonparametric inference methods given specification searches over different numbers of series terms in the nonparametric series regression model.  We provide methods of constructing uniform CIs and confidence bands by adjusting the conventional normal critical value to the critical value based on the supremum of the \textit{t}-statistics. The critical values can be constructed using simple Monte Carlo simulation or weighted bootstrap methods. Then, we provide an extension of the proposed CIs in the partially linear model setup. Finally, we investigate the  finite sample properties of the proposed methods and illustrate uniform CIs in an empirical example of Blomquist and Newey (2002).  

While beyond the scope of this paper, there are some potential directions to extend the results established here. First, investigating the coverage property of CIs with data-dependent $\widehat{K}$ using bias-corrected methods is of interest. In particular, it would be of interest to analyze the bias-corrected CI and confidence bands using cross-validation methods combined with the recent results established in Cattaneo, Farrell, and Feng (2019). Second, an extension of the current theory for quantile regression (e.g., Belloni, Chernozhukov, Chetverikov, and Fern\'{a}ndez-Val (2019)) or the  nonparametric IV setup would be desirable. In the NPIV setup, for example, one can consider pointwise CIs (or uniform confidence bands) that are uniform in pairs of $ (K_n, J_n) \in \mathcal{K}_n \times \mathcal{J}_n$ with an additional dimension of the instrument sieve and the number of instruments $J =J_n$. This is a difficult problem, and it would require a distinct theory to address the ill-posed inverse problem as well as two-dimensional choices. We leave these topics for future research.

\newpage
\section*{References}
\begin{description}
  \item[] \textsc{Andrews, D. W. K.} (1991a):  ``Asymptotic Normality of Series Estimators for Nonparametric and Semiparametric Regression Models,"\textit{Econometrica}, 59, 307-345.

  \item[] \textsc{Andrews, D. W. K.} (1991b):  ``Asymptotic Optimality of Generalized $C_L$, Cross-Validation, and Generalized Cross-Validation in Regression with Heteroskedastic Errors,"\textit{Journal of Econometrics}, 47, 359-377.

\item[] \textsc{Armstrong, T. B. and M. Koles\'{a}r} (2018):  ``A Simple Adjustment for Bandwidth Snooping,"\textit{Review of Economic Studies}, 85, 732-765.

   \item[] \textsc{Belloni, A., V. Chernozhukov, D. Chetverikov, and I. Fern\'{a}ndez-Val} (2019): ``Conditional quantile processes based on series or many regressors," \textit{Journal of Econometrics}, 213, 4-29. 

 \item[] \textsc{Belloni, A., V. Chernozhukov, D. Chetverikov, and K. Kato} (2015): ``Some New Asymptotic Theory for Least Squares Series: Pointwise and Uniform Results," \textit{Journal of Econometrics}, 186, 345-366.

  \item[] \textsc{Belloni, A., V. Chernozhukov, and C. Hansen} (2014): ``Inference on Treatment Effects after Selection among High-Dimensional Controls,"\textit{Review of Economic Studies}, 81, 608-650.

 \item[] \textsc{Blomquist, S. and W. K. Newey} (2002): ``Nonparametric Estimation with Nonlinear Budget Sets,'' \textit{Econometrica}, 70, 2455-2480.

 \item[] \textsc{Blundell, R. and T. E. MaCurdy} (1999): ``Labor Supply: A Review of Alternative Approaches," \textit{Handbook of Labor Economics}, In: O. Ashenfelter, D. Card (Eds.), vol. 3., Elsevier, Chapter 27.

  \item[] \textsc{Calonico, S., M. D. Cattaneo, and M. H. Farrell} (2018): ``On the Effect of Bias Estimation on Coverage Accuracy in Nonparametric Inference,"\textit{Journal of the American Statistical Association}, 113, 767-779.
  
    \item[] \textsc{Cattaneo, M. D. and M. H. Farrell} (2013): ``Optimal Convergence Rates, Bahadur Representation, and Asymptotic Normality of Partitioning Estimators," \textit{Journal of Econometrics},  174, 127-143.
  
  \item[] \textsc{Cattaneo, M. D., M. H. Farrell, and Y. Feng} (2019): ``Large Sample Properties of Partitioning-Based Series Estimators," \textit{Annals of Statistics},  forthcoming.
  
  \item[] \textsc{Cattaneo, M. D., M. Jansson, and W. K. Newey} (2018a): ``Alternative Asymptotics and the Partially Linear Model with Many Regressors," \textit{Econometric Theory}, 34, 277-301.

\item[] \textsc{Cattaneo, M. D., M. Jansson, and W. K. Newey} (2018b): ``Inference in Linear Regression Models with Many Covariates and Heteroscedasticity," \textit{Journal of the American Statistical Association}, 113, 1350-1361.

  \item[] \textsc{Chao, J. C., N. R. Swanson, J. A. Hausman, W. K. Newey, and T. Woutersen} (2012): ``Asymptotic Distribution of JIVE in a Heteroskedastic IV Regression with Many Instruments,"\textit{Econometric Theory}, 28, 42-86.

  \item[] \textsc{Chatterjee, S.} (2005): ``An error bound in the Sudakov-Fernique inequality,"arXiv:math/0510424

\item[] \textsc{Chen, X. } (2007): ``Large Sample Sieve Estimation of Semi-nonparametric Models," \textit{Handbook of Econometrics},
In: J.J. Heckman, E. Leamer (Eds.), vol. 6B., Elsevier, Chapter 76.

 \item[] \textsc{Chen, X.  and T. Christensen} (2015): ``Optimal Uniform Convergence Rates and Asymptotic Normality for Series Estimators Under Weak Dependence and Weak Conditions," \textit{Journal of Econometrics}, 188, 447-465.

  \item[] \textsc{Chen, X. and T. Christensen} (2018): ``Optimal Sup-norm Rates and Uniform Inference on Nonlinear Functionals of Nonparametric IV Regression", \textit{Quantitative Economics}, 9(1), 39-85.
  
\item[] \textsc{Chen, X. and Z. Liao} (2014): ``Sieve M inference on irregular parameters," \textit{Journal of Econometrics},182, 70-86.

\item[] \textsc{Chen, X., Z. Liao, and Y. Sun} (2014): ``Sieve inference on possibly misspecified semi-nonparametric time series models," \textit{Journal of Econometrics},178, 639-658.

\item[] \textsc{Chen, X. and X. Shen} (1998): ``Sieve extremum estimates for weakly dependent data," \textit{Econometrica}, 66 (2), 289-314.

\item[] \textsc{Chernozhukov V, D. Chetverikov, and K. Kato} (2014a): ``Gaussian approximation of suprema of empirical processes," \textit{The Annals of Statistics}, 42(4), 1564-1597.

\item[] \textsc{Chernozhukov V, D. Chetverikov, and K. Kato} (2014b): ``Anti-Concentration and Honest, Adaptive Confidence Bands,"\textit{The Annals of Statistics}, 42(5), 1787-1818.

\item[] \textsc{Chernozhukov V, D. Chetverikov, and K. Kato} (2016): ``Empirical and multiplier bootstraps for suprema of empirical processes of increasing complexity, and related Gaussian couplings,"\textit{Stochastic Processes and their Applications}, 126(12), 3632-3651.

\item[] \textsc{Donald, S. G. and W. K. Newey} (1994): ``Series Estimation of Semilinear Models,'' \textit{Journal of Multivariate Analysis}, 50, 30-40.

 \item[] \textsc{Eastwood, B. J. and  A.R. Gallant,} (1991): ``Adaptive Rules for Seminonparametric Estimators That Achieve Asymptotic Normality,"\textit{Econometric Theory}, 7, 307-340.

\item[] \textsc{Gin\'e, E.  and R. Nickl} (2010): ``Confidence bands in density estimation," \textit{The Annals of Statistics}, 38, 1122-1170.

\item[] \textsc{Gin\'e, E.  and R. Nickl} (2015): \textit{ Mathematical Foundations of Infinite-Dimensional Statistical Models}, Cambridge University Press.

\item[] \textsc{Hall, P. and J. Horowitz} (2013): ``A Simple Bootstrap Method for Constructing Nonparametric Confidence Bands for Functions," \textit{The Annals of Statistics}, 41, 1892-1921.

    \item[] \textsc{Hansen B. E.} (2015): ``The Integrated Mean Squared Error of Series Regression and a Rosenthal Hilbert-Space Inequality," \textit{Econometric Theory}, 31, 337-361.

 \item[] \textsc{Hansen, P.R.} (2005): ``A Test for Superior Predictive Ability,"\textit{Journal of Business and Economic Statistics}, 23, 365-380.

\item[] \textsc{H\"{a}rdle, W. and O. Linton} (1994): ``Applied Nonparametric Methods," \textit{Handbook of Econometrics},
In: R. F. Engle, D. F. McFadden (Eds.), vol. 4., Elsevier, Chapter 38.

\item[] \textsc{Hausman, J. A.} (1985): ``The Econometrics of Nonlinear Budget Sets", \textit{Econometrica}, 53, 1255-1282.

\item[] \textsc{Heckman, J. J., L. J. Lochner, and P. E. Todd} (2006): ``Earnings Functions, Rates of Return and Treatment Effects: The Mincer Equation and Beyond," \textit{Handbook of the Economics of Education},  In:  E. A. Hanushek, and F. Welch (Eds.), Vol. 1, Elsevier, Chapter 7.

 \item[] \textsc{Horowitz, J. L.} (2014): ``Adaptive Nonparametric Instrumental Variables Estimation: Empirical Choice of the Regularization Parameter," \textit{Journal of Econometrics}, 180, 158-173.

 \item[] \textsc{Horowitz, J. L. and S. Lee} (2012): ``Uniform Confidence Bands for Functions Estimated Nonparametrically with Instrumental Variables," \textit{Journal of Econometrics}, 168, 175-188.

  \item[] \textsc{Huang, J. Z.} (2003): ``Local Asymptotics for Polynomial Spline Regression," \textit{The Annals of Statistics}, 31, 1600-1635.
  
 \item[] \textsc{Kozbur, D.} (2018): ``Inference in Additively Separable Models With a High-Dimensional Set of Conditioning Variables," Working Paper, arXiv:1503.05436.

   \item[] \textsc{Leamer, E. E.} (1983): ``Let's Take the Con Out of Econometrics,"\textit{The American Economic Review}, 73, 31-43.

  \item[] \textsc{Lepski, O. V.} (1990): ``On a problem of adaptive estimation in Gaussian white noise,"\textit{Theory of Probability and its Applications}, 35, 454-466.

\item[] \textsc{Li, K. C.} (1987): ``Asymptotic Optimality for $C_p$, $C_L$, Cross-Validation and Generalized Cross-Validation: Discrete Index Set," \textit{The Annals of Statistics}, 15, 958-975.

\item[] \textsc{Li, Qi, and J. S. Racine} (2007): \textit{Nonparametric Econometrics: Theory and Practice}, Princeton University Press.

\item[] \textsc{Linton, O.} (1995): ``Second order approximation in the partialy linear regression model," \textit{Econometrica}, 63(5), 1079-1112.

\item[] \textsc{Newey, W. K.} (1994a): ``Series Estimation of Regression Functionals,'' \textit{Econometric Theory}, 10, 1-28.

\item[] \textsc{Newey, W. K.} (1994b): ``The Asymptotic Variance of Semiparametric Estimators,'' \textit{Econometrica}, 62, 1349-1382.

  \item[] \textsc{Newey, W. K.} (1997): ``Convergence Rates and Asymptotic Normality for Series Estimators,''\textit{Journal of Econometrics}, 79, 147-168.

   \item[] \textsc{Newey, W.  K.} (2013): ``Nonparametric Instrumental Variables Estimation,''\textit{American Economic Review: Papers \& Proceedings}, 103, 550-556.

\item[] \textsc{Newey, W. K. and J. L. Powell} (2003): ``Instrumental Variable Estimation of Nonparametric Models,''\textit{Econometrica}, 71, 1565-1578.

\item[] \textsc{Newey, W. K. and J. L. Powell, F. Vella} (1999): ``Nonparametric Estimation of Triangular Simultaneous Equations Models,''\textit{Econometrica}, 67, 565-603.

  \item[] \textsc{Robinson, P. M.} (1988): ``Root-N-Consistent Semiparametric Regression,''\textit{Econometrica}, 56(4), 931-954.

  \item[] \textsc{Romano, J. P. and M. Wolf} (2005): ``Stepwise Multiple Testing as Formalized Data Snooping,''\textit{Econometrica}, 73, 1237-1282.

  \item[] \textsc{Schennach, S. M.} (2015): ``A bias bound approach to nonparametric inference," \textit{CEMMAP working paper CWP71/15}.

    \item[] \textsc{Van Der Vaart, A. W. and J. A. Wellner} (1996):\textit{ Weak Convergence and Empirical Processes}, Springer.
    
  \item[] \textsc{White, H.} (2000): ``A Reality Check for Data Snooping,''\textit{Econometrica}, 68, 1097-1126.

  \item[] \textsc{Zhou, S., X. Shen, and D.A. Wolfe} (1998): ``Local Asymptotics for Regression Splines and Confidence Regions," \textit{The Annals of Statistics}, 26, 1760-1782.

\end{description}

\newpage
\begin{appendices}
\section{Proofs}
\label{sec:proof}

\subsection{Preliminaries and Useful Lemmas}

We define additional notations for the empirical process theory used in the proof of Theorem \ref{thm:uniform}. Given measurable space $(S, \mathcal{S})$, let $\mathcal{F}$ as a class of measurable functions $f: \mathcal{S} \rightarrow \mathbb{R}$. For any probability measure $Q$ on $(S, \mathcal{S})$,  we define $N(\epsilon, \mathcal{F}, L_2(Q))$ as covering numbers, which is the minimal number of the $L_{2}(Q)$ balls of radius $\epsilon$ to cover $\mathcal{F}$ with $L_2(Q)$ norms $ || f ||_{Q,2} = (\int |f|^2 dQ)^{1/2}$.  The uniform entropy numbers relative to the $L_2(Q)$ norms are defined as $\sup_{Q} \log N(\epsilon ||F||_{Q,2}, \mathcal{F}, L_2(Q)) $ where  the supremum is over all discrete probability measures with an envelope function $F$.  We define $\mathcal{F}$ as a \textit{VC type} with envelope $F$ if there are constants $A, v > 0$ such that $\sup_{Q} N(\epsilon ||F||_{Q,2}, \mathcal{F}, L_2(Q))  \leq (A/\epsilon)^{v}$ for all $0 < \epsilon \leq 1$.

Let the data $z_i = (\varepsilon_i, x_i) $ be i.i.d. random vectors defined on the probability space $(\mathcal{Z} = \mathcal{E} \times \mathcal{X}, \mathcal{A}, P)$ with common probability distribution $P \equiv P_{\varepsilon, x}$. We think of $(\varepsilon_1, x_1), \cdots (\varepsilon_n, x_n)$ as the coordinates of the infinite product probability space. We avoid discussing nonmeasurability issues and outer expectations (for the related issues, see van der Vaart and Wellner (1996)). Throughout the proofs,  we denote $c, C >0$ as universal constants that do not depend on $n$.

For any sequence $\{ K = K_n  : n \geq 1 \} \in \prod_{n=1}^{\infty} \mathcal{K}_n$ under Assumption \ref{assump:sets_different}, we first define the orthonormalized vector of basis functions
\begin{equation*}
\tilde{P} (K,x) \equiv Q_K^{-1/2} P (K, x) = E[P_{Ki} P_{Ki}']^{-1/2} P (K, x), \ \tilde{P}_{Ki} = \tilde{P}(K, x_i), \ \tilde{P}^K = [\tilde{P}_{K1}, \cdots, \tilde{P}_{Kn}]'.
\end{equation*}
We observe that
\begin{equation*}
\widehat{g}_n(K, x) = \tilde{P} (K, x)' (\tilde{P}^{K'} \tilde{P}^{K})^{-1} \tilde{P}^{K'} Y, \ \ V_n(K, x) = \tilde{P} (K,x)' \tilde{\Omega}_K  \tilde{P} (K,x), \ \  \tilde{\Omega}_K = E(\tilde{P}_{Ki} \tilde{P}_{Ki}' \varepsilon_{i}^2).
\end{equation*}

\noindent Without loss of generality, we may impose normalizations of $Q_{\overline{K}} = I_{\overline{K}}$ or $Q_{K}  = E(P_{Ki} P_{Ki}') = I_{K}$ uniformly over $K \in \mathcal{K}_n$, since $\widehat{g}_n(K, x)$ is invariant to nonsingular linear transformations of $P(K, x)$.  However, we shall treat $Q_K$ as unknown and deal with the non-orthonormalized series terms. Next, we re-define pseudo true value $\beta_K$, with an abuse of notation, using orthonormalized series terms $\tilde{P}_{Ki}$. That is, $y_i = \tilde{P}_{Ki}' \beta_K + \varepsilon_{Ki}, E [\tilde{P}_{Ki} \varepsilon_{Ki}] = 0$ where  $\varepsilon_{Ki} = r_{Ki} + \varepsilon_i$, $r_n (K, x) = g_0(x) - \tilde{P} (K, x)'\beta_K,  r_{Ki} = r_n(K, x_i)$, and $r_K  \equiv (r_{K1}, \cdots r_{K n})'$. We also define $\widehat{Q}_K \equiv \frac{1}{n} \tilde{P}^{K'} \tilde{P}^{K}$, $\underline{\sigma}^2 \equiv \inf_{x} E[\varepsilon_i^2 |x_i=x], \bar{\sigma}^2 \equiv \sup_{x} E[\varepsilon_i^2 |x_i=x]$.

We first provide useful lemmas which will be used in the proof of Theorem \ref{thm:asymp-cov} and \ref{thm:uniform}. The versions of proof of Lemmas \ref{lemma:bounds} and \ref{lemma:bounds-uniform} with $\mathcal{K}_n = \{ K \}$ are available in the literature, such as  Belloni et al. (2015) and Chen and Christensen (2015), among many others. The maximal inequalities are used in the proof of Lemmas \ref{lemma:bounds} and \ref{lemma:bounds-uniform} to bound the remainder terms in the linearization of the \textit{t}-statistics. Also note that different rate conditions of $K$ such as those in Newey (1997)  can be used here but lead to different bounds.  We provide the proofs of Lemma \ref{lemma:bounds} and \ref{lemma:bounds-uniform} in the Online Supplementary Material (Section B).

\begin{lemma}\label{lemma:bounds} Suppose that Assumptions \ref{assump:sets_different}, \ref{assump:model}, and \ref{assump:regularity} hold, then $|| \widehat{Q}_K - I_K || = O_p(\sqrt{\lambda_K^2 \zeta_K^2\log K/n})$ for any $K \in \mathcal{K}_n$ and the following holds
\begin{align}
\label{R1}&  \max_{K \in \mathcal{K}_n} |R_1(K, x) | = O_p(\max_{K \in \mathcal{K}_n}\sqrt{\frac{\lambda_K^2 \zeta_K^2 \log K \log p}{n}} (1+ \ell_K c_K \sqrt{K})),\\
\label{R2} &\max_{K \in \mathcal{K}_n} |R_2(K, x) |  =  O_p(\max_{K \in \mathcal{K}_n}(\ell_{K} c_{K}) \sqrt{\log p}),
\end{align}
where $R_1(K, x) \equiv \sqrt{ \frac{1}{n V_n(K, x)}} \tilde{P} (K, x)' ( \widehat{Q}_K^{-1} - I_K) \tilde{P}^{K'} (\varepsilon+ r_K), R_2(K,x)  \equiv \sqrt{ \frac{1}{n V_n(K,x)}} \tilde{P}(K, x)' \tilde{P}^{K'} r_K$.
\end{lemma}

\begin{lemma}\label{lemma:bounds-uniform} Suppose that Assumptions \ref{assump:sets_different}, \ref{assump:model} and \ref{assump:regularity-uniform} hold, then the following holds
\begin{align}
\label{R1-uniform}&  \sup_{ K \in \mathcal{K}_n, x \in \mathcal{X}} |R_1(K, x) | = O_p(\max_{K \in \mathcal{K}_n}\sqrt{\frac{\lambda_K^2 \zeta_K^2 \log K \log n}{n}} (n^{1/q}+ \ell_K c_K \sqrt{K})),\\
\label{R2-uniform} &\sup_{ K \in \mathcal{K}_n, x \in \mathcal{X}} |R_2(K, x) |  =  O_p(\max_{K \in \mathcal{K}_n}(\ell_{K} c_{K}) \sqrt{\log n}),
\end{align}
where $R_1(K, x), R_2(K,x)$ are defined in Lemma \ref{lemma:bounds}.
\end{lemma}

 \subsection{Proofs of the Main Results}

\subsubsection{Proof of Theorem \ref{thm:asymp-cov}}
\begin{proof}

For any $K \in \mathcal{K}_n$, we first consider the decomposition of the \textit{t}-statistic in \eqref{eq:t-decompose} with the known variance $V_n(K,x)$,
\begin{align*}
T_n (K, x) & = \sqrt{ \frac{n}{V_n(K, x)}} \tilde{P}(K, x)'(\widehat{\beta}_{K} -\beta_{K}) -  \sqrt{\frac{n}{V_n(K, x)}} r_n(K, x)  \\
 & = t_n(K, x)+ R_1(K,x) + R_2(K,x)  + \nu_n (K,x)
\end{align*}
\noindent where $t_n(K,x) = n^{-1/2} \sum_{i=1}^{n} \frac{\tilde{P} (K, x)'\tilde{P}_{K i} \varepsilon_i}{V_n(K, x)^{1/2}}$, $R_1(K,x), R_2(K,x)$ are defined in Lemma \ref{lemma:bounds}, and $\nu_n (K,x) =-  \sqrt{n} V_n(K,x)^{-1/2} r_n(K,x)$. 
Define 
\begin{equation*}
t_n \equiv (t_n(K_1,x), \cdots, t_n(K_p,x))^{\prime} =\frac{1}{\sqrt{n}} \sum_{i=1}^{n} \xi_i
\end{equation*}
where $\xi_i = (\xi_{i1}, \xi_{i2}, \cdots, \xi_{ip})^{\prime} \in \mathbb{R}^{p}$ with $\xi_{ij} =  \frac{\tilde{P} (K_j, x)'\tilde{P}_{K_j i} \varepsilon_i}{V_n(K_j, x)^{1/2}}$ and $p = |\mathcal{K}_n|$. Note that $E[\xi_{ij}] = 0$ and $E[|\xi_{ij}|^3] \lesssim E[|\tilde{P} (K_j, x)'\tilde{P}_{K_j i}/V_n(K_j, x)^{1/2}|^3] \sup_{x} E[|\varepsilon_i|^3 | x_i = x] \lesssim \max_{K} \zeta_K $ for all $1\leq i \leq n, 1 \leq j \leq p$. By Lemma A.2 in the Online Supplementary Material, for any $\delta >0$, there exists a random variable $\max_{1\leq j\leq p} \sum_{i=1}^{n} Z_{ij}$ with independent random vectors $\{ Z_i \}_{i=1}^{n} \in \mathbb{R}^{p}$, $Z_{i} \sim N(0, \frac{1}{n} E[\xi_{i} \xi_{i}^{\prime}]), 1\leq i \leq n$, such that
\begin{equation*}
P(|\max_{1\leq j\leq p}| t_n(K_j,x)| - \max_{1\leq j\leq p} \sum_{i=1}^{n} |Z_{ij}||  > 16 \delta) \lesssim \frac{ \log(p\vee n)}{ \delta^{2}} D_1 + \frac{ \log^{2}(p\vee n)}{\delta^3 n^{3/2}} (D_2 + D_3) + \frac{\log n}{n}
\end{equation*}
where $D_1 = E\big[ \max_{1\leq j, l \leq p} |\frac{1}{n} \sum_{i=1}^{n} (\xi_{ij} \xi_{il} - E[\xi_{ij} \xi_{il}] ) |\big], D_2 = E\big[ \max_{1\leq j \leq p} \sum_{i=1}^{n} | \xi_{ij} |^{3} \big]$, and $D_3 = \sum_{i=1}^{n} E\big[ \max_{1\leq j \leq p} |\xi_{ij}|^{3} 1 \big( \max_{1\leq j \leq p} |\xi_{ij}| > \delta \sqrt{n}/\log (p\vee n) \big) \big].$

First consider the case (a) in Assumption \ref{assump:regularity}(\rmnum{2}). Combining bounds for $D_1, D_2, D_3$ in Lemma B.1 in the Online Supplementary Material gives, for any $\delta>0$, 
\begin{eqnarray*}
&& \hspace{-0.5cm}P(|\max_{1\leq j\leq p}| t_n(K_j,x)| - \max_{1\leq j\leq p} \sum_{i=1}^{n} |Z_{ij}||  > 16 \delta) \\
&&\lesssim \frac{ \log(p\vee n)}{ \delta^{2}} \big[  (\frac{ (\max_{K} \zeta_K)^2 \log p}{n})^{1/2} + \frac{(\max_{K} \zeta_K)^{2}\log p }{n^{1 - 2/q}} \big]  \\
&&+ \frac{ \log^{2}(p\vee n)}{\delta^3 } \big[ (\frac{ (\max_{K} \zeta_K)^2}{n})^{1/2} + \frac{(\max_{K} \zeta_K)^{3} \log p}{n^{3/2 - 3/q}}  \big] + \frac{ \log^{q-1}(p\vee n)}{\delta^{q}} \frac{(\max_{K} \zeta_K)^{q} }{n^{q/2-1}} + \frac{\log n}{n}. 
\end{eqnarray*}
For $\gamma >0$, by setting 
\begin{eqnarray*}
\delta  &=& \gamma^{-1/3} \big( \frac{(\max_{K} \zeta_K)^{2} \log^{4} (p \vee n) }{n} \big)^{1/6} +\gamma^{-1/2}  \big( \frac{(\max_{K} \zeta_K)^{2}\log (p \vee n) \log p}{n^{1 - 2/q} }  \big)^{1/2}\\
&&+ \gamma^{-1/3}  \big( \frac{(\max_{K} \zeta_K)^{3}\log^{2} (p \vee n) \log p}{n^{3/2 - 3/q} }  \big)^{1/3}, 
\end{eqnarray*}
we have 
\begin{equation*}
P(|\max_{1\leq j\leq p}| t_n(K_j,x)| - \max_{1\leq j\leq p} \sum_{i=1}^{n} |Z_{ij}||  >  C_1 \delta ) \leq C_2 (\gamma + \frac{\log n}{n})
\end{equation*}
where $C_1, C_2$ are positive constants that depend only on $q$. If we take $\gamma = \gamma_n \rightarrow 0$ sufficiently slowly, e.g., $\gamma = \log (p \vee n)^{-1/2}$, then the above implies there exists $\max_{1\leq j\leq p} \sum_{i=1}^{n} Z_{ij}$ such that 
\begin{equation*}
|\max_{1\leq j\leq p}| t_n(K_j,x)| - \max_{1\leq j\leq p} \sum_{i=1}^{n} |Z_{ij}|| = o_p(\big( \frac{(\max_{K} \zeta_K)^{2} \log^{5} (p \vee n) }{n} \big)^{1/6}  + \frac{(\max_{K} \zeta_K) \log^{3/4} (p \vee n) \log^{1/2} p }{n^{1/2-1/q}}).
\end{equation*}

\noindent Next, consider the case (b) in Assumption \ref{assump:regularity}(\rmnum{2}). For any $\delta>0$, 
\begin{eqnarray*}
&& \hspace{-0.5cm}P(|\max_{1\leq j\leq p}| t_n(K_j,x)| - \max_{1\leq j\leq p} \sum_{i=1}^{n} |Z_{ij}||  > 16 \delta) \\
&&\lesssim \frac{ \log(p\vee n)}{ \delta^{2}} \big[  (\frac{ (\max_{K} \zeta_K)^2 \log p}{n})^{1/2} + \frac{(\max_{K} \zeta_K)^{2} \log^{2} (pn) \log p }{n} \big]  \\
&&+ \frac{ \log^{2}(p\vee n)}{\delta^3 } \big[ (\frac{ (\max_{K} \zeta_K)^2}{n})^{1/2} + \frac{(\max_{K} \zeta_K)^{3} \log^{3}(pn)\log p}{n^{3/2}}  \big] \\
&& +  \frac{ \log^{2}(p\vee n)}{\delta^3 } \big[ \frac{1}{n^{1/2}}(\frac{\delta^3 n^{3/2}}{\log^3 (p\vee n)} + (\max_{K} \zeta_K)^3 \log^3 p)\exp( - \frac{\delta \sqrt{n}}{C \max_{K} \zeta_K\log p\log (p\vee n) } ) \big] + \frac{\log n}{n}
\end{eqnarray*}
by Lemma B.1 in the Online Supplementary Material. Similarly, by setting
\[
\delta = \max \{ \gamma^{-1/3} (\max_{K} \zeta_K)^{2} \log^{4} (p \vee n)/n)^{1/6}, 2 C ((\max_{K} \zeta_K)^{2} \log^{4} (p \vee n) \log^2 p/n)^{1/2} \}
\]
we have, for  $\gamma>0$, 
\begin{equation*}
P(|\max_{1\leq j\leq p}| t_n(K_j,x)| - \max_{1\leq j\leq p} \sum_{i=1}^{n} |Z_{ij}||  >  C_1 \delta ) \leq C_2 (\gamma + \frac{\log n}{n})
\end{equation*}
where $C_1, C_2$ are universal constants which do not depend on $n$. Here we use  $\frac{\delta\sqrt{n}}{C\max_{K} \zeta_K\log p\log (p\vee n)} \geq 2 \log (p \vee n )$. By taking $\gamma = \log (p \vee n)^{-1/2}$, there exists $\max_{1\leq j\leq p} \sum_{i=1}^{n} Z_{ij}$ such that 
\begin{equation*}
|\max_{1\leq j\leq p}| t_n(K_j,x)| - \max_{1\leq j\leq p} \sum_{i=1}^{n} |Z_{ij}|| = o_p(\big( \frac{(\max_{K} \zeta_K)^{2} \log^{5} (p \vee n) }{n} \big)^{1/6}  + \frac{(\max_{K} \zeta_K)^{2} \log^{4} (p \vee n) \log^2 p }{n} \big)^{1/2}).
\end{equation*}

In either case (a) or (b), the above coupling inequality shows that there exists a sequence of random variables $\max_{ 1 \leq j \leq p} \sum_{i=1}^{n} |Z_{ij}|$ such that $\big|\max_{K \in \mathcal{K}_{n}} |t_n(K,  x)| - \max_{ 1 \leq j \leq p} \sum_{i=1}^{n} |Z_{ij}|\big| = o_p(a_{n})$, $a_{n} =1/(\log p)^{1/2}$ under the rate conditions imposed in Theorem \ref{thm:asymp-cov}. Furthermore, 
\begin{align}
\nonumber\big|   \max_{1 \leq j \leq p} |T_n(K_j, x)|  - \max_{1 \leq j \leq p} |t_n(K_j, x)|\big| &\leq \max_{1 \leq j \leq p} |T_n(K_j, x)- t_n(K_j, x)| \leq \max_{1 \leq j \leq p} |R_1(K_j,x)| \\
\label{eq:coupling-Tt}& \quad +  \max_{1 \leq j \leq p} |R_2(K_j,x)| +  \max_{1 \leq j \leq p} |\nu_n(K_j,x)| = o_p(a_n)
\end{align}
with $a_{n} =1/(\log p)^{1/2}$ by Lemma \ref{lemma:bounds} and the assumption imposed in Theorem  \ref{thm:asymp-cov}. We also have 
\begin{align}
\nonumber\big|   & \max_{1 \leq j \leq p} |T_n(K_j, x)|  - \max_{1 \leq j \leq p} |\widehat{T}_n(K_j, x)|\big| \leq \max_{1 \leq j \leq p} |T_n(K_j, x)- \widehat{T}_n(K_j, x)| \\
\label{eq:coupling-TT_hat}\leq & \max_{1 \leq j \leq p} |T_n(K_j, x)| \max_{1 \leq j \leq p} |1 - \frac{V_n(K_{j},x)^{1/2}}{\widehat{V}_n(K_{j},x)^{1/2}}| = o_p(a_n)
\end{align}
where we use Lemma \ref{lemma:bounds} and $ \max_{1 \leq j \leq p} |t_n(K_j,x)| \lesssim_{P} \sqrt{\log p}$ by the maximal inequality (e.g., Lemma A.4 in the Online Supplementary Material) and Assumption \ref{assump:regularity}(\rmnum{3}) with $a_{n} =1/(\log p)^{1/2}$. Combining \eqref{eq:coupling-Tt} and \eqref{eq:coupling-TT_hat} gives $\big|   \max_{1 \leq j \leq p} |\widehat{T}_n(K_j, x)|  - \max_{ 1 \leq j \leq p} \sum_{i=1}^{n} |Z_{ij}|\big|  = o_p(a_n)$ with $a_n = 1/(\log p)^{1/2} $. Then, there exists some sequence of positive constant $\delta_n$ such that $\delta_n = o(1)$  and $P(\big|   \max_{1 \leq j \leq p} |\widehat{T}_n(K_j, x)|  -\max_{ 1 \leq j \leq p} \sum_{i=1}^{n} |Z_{ij}|  \big|> a_n \delta_n ) = o(1)$. 

For any $u \in \mathbb{R}$, we have
\begin{align*}
&P(\max_{1 \leq j \leq p} |\widehat{T}_n(K_j, x)| \leq u) \\
&\leq P(\{ \max_{1 \leq j \leq p} |\widehat{T}_n(K_j, x)| \leq u \}  \cap \{ \big|   \max_{1 \leq j \leq p} |\widehat{T}_n(K_j, x)|  - \max_{ 1 \leq j \leq p} \sum_{i=1}^{n} |Z_{ij}|\big| \leq a_n \delta_n \}) \\
&\quad +  P (\big|   \max_{1 \leq j \leq p} |\widehat{T}_n(K_j, x)|  - \max_{ 1 \leq j \leq p} \sum_{i=1}^{n} |Z_{ij}|\big| > a_n \delta_n) \\
&\leq P(\max_{ 1 \leq j \leq p} \sum_{i=1}^{n} |Z_{ij}| \leq u + a_n\delta_n) + o(1) \leq P(\max_{ 1 \leq j \leq p} \sum_{i=1}^{n} |Z_{ij}| \leq u ) + a_n\delta_n E[\max_{ 1 \leq j \leq p} \sum_{i=1}^{n} |Z_{ij}| ]+ o(1) 
\end{align*}
where the last inequality uses anti-concentration inequality (Lemma A.8 in the Online Supplementary Material). The reverse inequality holds with a similar argument above, and thus 
\begin{equation*}
\sup_{u \in \mathbb{R}} \big| P(\max_{1 \leq j \leq p} |\widehat{T}_n(K,  x) | \leq u) - P(\max_{ 1 \leq j \leq p} \sum_{i=1}^{n} |Z_{ij}| \leq u)  \big| = a_n\delta_n E[\max_{ 1 \leq j \leq p} \sum_{i=1}^{n} |Z_{ij}| ]+ o(1) 
 =  o(1)
\end{equation*}
where we use $E[\max_{ 1 \leq j \leq p} \sum_{i=1}^{n} |Z_{ij}| ] \lesssim \sqrt{\log p}$ by Gaussian maximal inequality and $a_n =(\log p)^{-1/2} $. Using the same arguments above,  $\big|   \max_{ 1 \leq j \leq p} \sum_{i=1}^{n} |Z_{ij}|  -\max_{ 1 \leq j \leq p} \sum_{i=1}^{n} |\widehat{Z}_{ij}|\big| = o_p(a_{n})$ by Sudakov-Fernique type bound (e.g., Chatterjee (2005)) and  Assumption \ref{assump:regularity}(\rmnum{3}), we have $\sup_{u \in \mathbb{R}} \big| P(\max_{ 1 \leq j \leq p}  |\widehat{Z}_{ij}| | \leq u) - P(\max_{ 1 \leq j \leq p} \sum_{i=1}^{n} |Z_{ij}| \leq u) |= o(1)$. Therefore, the following holds by the triangle inequality,
\begin{equation*}
\sup_{u \in \mathbb{R}} \big| P(\max_{1 \leq j \leq p} |\widehat{T}_n(K,  x) | \leq u) - P(\max_{ 1 \leq j \leq p} \sum_{i=1}^{n} |\widehat{Z}_{ij}| \leq u)  \big| =  o(1), 
\end{equation*}
and then we conclude 
\begin{equation*}
P(\max_{K \in \mathcal{K}_{n}} |\widehat{T}_n(K,  x) |  \leq \widehat{c}_{1-\alpha}(x) )  = 1-\alpha + o(1),
\end{equation*}
with a critical value $\widehat{c}_{1-\alpha}(x) $ given in \eqref{def:cv-mcsup}, and the coverage result \eqref{result:coverage-uniformci} follows. 

Finally, we will show \eqref{result:coverage-pmsci}. For $\widehat{K} \in \mathcal{K}_n$, observe that
\begin{equation}\label{eq:TnKhat_triangle}
|\widehat{T}_n(\widehat{K},x)| \leq (|t_n(\widehat{K},x)| +  |R_1(\widehat{K},x)| +  |R_2(\widehat{K},x)| + |\nu_n(\widehat{K},x)|) |\frac{V_n(\widehat{K},x)^{1/2}}{\widehat{V}_n(\widehat{K},x)^{1/2}}|
\end{equation}
by the triangle inequality. Then, 
\begin{align}
\nonumber&P( g_0(x) \in  [ \widehat{g}_n(\widehat{K},x) \pm \widehat{c}_{1-\alpha} (x) \sqrt{\widehat{V}_n(\widehat{K}, x)/n} ]) \\
\nonumber&\geq P(|t_n(\widehat{K},x)| +  |R_1(\widehat{K},x)| +  |R_2(\widehat{K},x)| + |\nu_n(\widehat{K},x)| \leq \widehat{c}_{1-\alpha}(x)|\frac{\widehat{V}_n(\widehat{K},x)^{1/2}}{V_n(\widehat{K},x)^{1/2}}| )\\
\label{eq:pmsci-proof1}&\geq P(|t_n(\widehat{K},x)| +  |R_1(\widehat{K},x)| +  |R_2(\widehat{K},x)| + |\nu_n(\widehat{K},x)| \leq \widehat{c}_{1-\alpha}(x) (1-a_n^2 \delta_{1n})) - \epsilon_{1n}\\
\label{eq:pmsci-proof2}&\geq P(|t_n(\widehat{K},x)| \leq \widehat{c}_{1-\alpha}(x) (1-a_n^2 \delta_{1n}) - a_n \delta_{2n} - a_n \delta_{3n}) - \epsilon_{1n} -  \epsilon_{2n} -  \epsilon_{3n}\\
\label{eq:pmsci-proof3}&\geq P(\max_{K \in \mathcal{K}_n} |t_n(K,x)| \leq \widehat{c}_{1-\alpha}(x) (1-a_n^2 \delta_{1n}) - a_n \delta_{2n} - a_n\delta_{3n}) - \epsilon_{1n} -  \epsilon_{2n} -  \epsilon_{3n}\\
\label{eq:pmsci-proof4}&\geq P(\max_{ 1 \leq j \leq p} \sum_{i=1}^{n} |\widehat{Z}_{ij}| \leq \widehat{c}_{1-\alpha}(x) - \tilde{\delta}_n) - \tilde{\epsilon}_n\\
\label{eq:pmsci-proof5}& \geq 1-\alpha - \sup_{u} P( | \max_{ 1 \leq j \leq p} \sum_{i=1}^{n} |\widehat{Z}_{ij}| - u| \leq \tilde{\delta}_n) -   \tilde{\epsilon}_n \ \geq 1-\alpha - o(1).
\end{align}
The first inequality follows by \eqref{eq:TnKhat_triangle}, and \eqref{eq:pmsci-proof1} holds by Assumption \ref{assump:regularity}(\rmnum{3}) with some sequence of positive constant $\delta_{1n} =o(1), \epsilon_{1n} = o(1)$ and  \eqref{eq:pmsci-proof2} follows by $|R_1(\widehat{K},x)| +  |R_2(\widehat{K},x)| = o_p(a_n)$ from Lemma \ref{lemma:bounds}  and the assumption $|\frac{\sqrt{n}r_n(\widehat{K},x)}{V_{n}(\widehat{K},x)^{1/2}} | = o(a_n)$ with $a_n =1/(\log p)^{1/2} $ and some sequences of constants $\delta_{2n} =o(1), \epsilon_{2n} = o(1), \delta_{3n} =o(1), \epsilon_{3n} = o(1)$.  \eqref{eq:pmsci-proof3} follows by $|t_n(\widehat{K},x)| \leq \max_{K \in \mathcal{K}_n} |t_n(K,x)|$,  and  \eqref{eq:pmsci-proof4} holds by $| \max_{K \in \mathcal{K}_n} |t_n(K,x)| - \max_{ 1 \leq j \leq p} \sum_{i=1}^{n} |\widehat{Z}_{ij}|| = o_p(a_n)$ with some sequences  $\delta_{4n} =o(1), \epsilon_{4n} = o(1)$ and defining $\tilde{\delta}_n = \widehat{c}_{1-\alpha}(x) a_n^2 \delta_{1n} + a_n \delta_{2n} + a_n \delta_{3n} + a_n\delta_{4n}, \tilde{\epsilon}_n = \epsilon_{1n} + \epsilon_{2n} + \epsilon_{3n} + \epsilon_{4n}$. Finally, \eqref{eq:pmsci-proof5} holds by Lemma A.8, $E[\max_{ 1 \leq j \leq p} \sum_{i=1}^{n} |\widehat{Z}_{ij}| ] \lesssim \sqrt{\log p}$ and $ \tilde{\delta}_n \sqrt{\log p}= o(1)$ since $ \widehat{c}_{1-\alpha}(x)   \lesssim \sqrt{\log p} $  by Lemma A.15. This completes the proof. 
\end{proof}

\subsubsection{Proof of Theorem \ref{thm:uniform} }
\begin{proof}
Similar to the proof of Theorem \ref{thm:asymp-cov}, we have the following linearization of the \textit{t}-statistics uniformly in $(K, x) \in \mathcal{K}_n \times \mathcal{X}$, 
\begin{equation*} 
T_n(K, x)  =t_n(K,x)  +  \nu_n(K,x)+  R_n(K,x),
\end{equation*}
where $t_n(K,x) = n^{-1/2} \sum_{i=1}^{n} \tilde{P}(K, x)' \tilde{P}_{K i } \varepsilon_i/V_n(K,x)^{1/2}$ and $R_n(K,x) = R_1(K,x) + R_2(K,x)$. Define $f_{n,K,x} : (\mathcal{E} \times \mathcal{X}) \mapsto \mathbb{R}$ for given $n \geq 1$, $K \in \mathcal{K}_{n}, x \in \mathcal{X}$, 
\begin{equation}
f_{n,K, x} (\varepsilon, t) = \frac{\tilde{P} (K, x)' \tilde{P} (K, t) \varepsilon}{V_{n}(K,x)^{1/2}}, (\varepsilon, t) \in \mathcal{E} \times \mathcal{X}.
\end{equation}
and consider the class of measurable functions $\mathcal{F}_n  = \{ f_{n,K,x} : (K,x) \in \mathcal{K}_n \times \mathcal{X} \}$. Then, we consider the following empirical process:
\[
\Big\{ t_n(K,x): (K,x) \in \mathcal{K}_n \times \mathcal{X} \Big\} = \Big\{ n^{-1/2} \sum_{i=1}^{n} f_{n,K,x} (\varepsilon_i, x_i): (K,x) \in \mathcal{K}_n \times \mathcal{X}\Big\}
\]
which is indexed by classes of functions $\mathcal{F}_n $.  Define $\alpha(K, x) \equiv \tilde{P}(K,x)/ V_n(K,x)^{1/2}  = \tilde{P}(K, x)/|| \Omega^{1/2}_K \tilde{P} (K, x) || $. Note that $|f_{n,K,x} (\varepsilon, t)| = | \alpha(K,x) ' \tilde{P} (K, t) \varepsilon| \leq C |\varepsilon| \max_K \zeta_{K}$ for any $(K, x)\in \mathcal{K}_n \times \mathcal{X}$. We define the envelope function $F_n(\varepsilon, t) \equiv  C |\varepsilon| \max_K \zeta_{K} \vee 1$.  By Assumption \ref{assump:regularity-uniform},  we have
\begin{align*}
&|f_{n,K,x} - f_{n, K^{\prime}, x^{\prime}}| = |\varepsilon| | \alpha(K,x)' \tilde{P} (K, t) -  \alpha(K^{\prime},x^{\prime})' \tilde{P} (K^{\prime}, t)|\\
& \leq |\varepsilon| | \big[ |\alpha(K,x)' \tilde{P} (K, t)  -  \alpha(K,x)' \tilde{P} (K^{\prime}, t)|  +| \alpha(K,x)' \tilde{P} (K^{\prime}, t)   - \alpha(K^{\prime},x)' \tilde{P} (K^{\prime}, t)|   \\
&\quad + |\alpha(K^{\prime},x)' \tilde{P} (K^{\prime}, t) -  \alpha(K^{\prime},x^{\prime})' \tilde{P} (K^{\prime}, t)|\big] \leq  |\varepsilon|  A \max_{K} \zeta_{K} L_n ( || x - x^{\prime}||  + | K - K^{\prime}| ) 
\end{align*}
for all $x, x^{\prime} \in \mathcal{X}, K, K^{\prime} \in \mathcal{K}_n$ where $L_n = \zeta^{L_1} \vee \zeta^{L_2}$. Therefore, the class of functions $\mathcal{F}_n  = \{ f_{n,K,x} : (K,x) \in \mathcal{K}_n \times \mathcal{X} \}$ is a VC type and there are constants $A, V>0$ such that 
\begin{equation*}
\sup_{Q} N(\epsilon ||F_n||_{L^2(Q)}, \mathcal{F}_n, L^2(Q)) \leq (A L_n/\epsilon)^V , 0 < \forall \epsilon \leq 1
\end{equation*}
for each $n$. Then, using Theorem 2.1 (Lemma A.9 in the Online Supplementary Material)  in Chernozhukov et al. (2016) with $B(f)=0$, there exists a tight Gaussian process $G_n(f)$ in $\ell^{\infty}( \mathcal{F}_{n})$ and $Z_n(K,x) = G_n(f_{n, K,x})$ in $\ell^{\infty}( \mathcal{K}_{n} \times \mathcal{X})$ with zero mean and covariance function \eqref{def:covariance-uniform}, $E[ G_n(f) G_n(f^{\prime})] = Cov(f_{n, K, x}(\varepsilon_i,x_i), f_{n, K^{\prime}, x^{\prime}}^{\prime}(\varepsilon_i,x_i))$ and a sequence of random variables $\widetilde{Z} \equiv \sup_{(K, x) \in \mathcal{K}_{n} \times \mathcal{X}} |Z_n (K,x)|$ such that, for every $\gamma  \in (0,1)$, 
\begin{equation}
P(|\sup_{(K, x) \in \mathcal{K}_{n} \times \mathcal{X}} |t_n (K,x)|  - \widetilde{Z}| > C_1 \delta_{1n} ) \leq C_2 (\gamma + n^{-1})
\end{equation}
where $C_{1}, C_{2}$ are positive constants that depend only on $q$, and 
\[
\delta_{1n} = \gamma^{-1/q} n^{-1/2 + 1/q} \max_{K} \zeta_K \log n + \gamma^{-1/3}n^{-1/6} (\max_{K} \zeta_K)^{1/3} \log^{2/3} n
\]
by Assumption \ref{assump:regularity-uniform}(\rmnum{3}) and assuming $\log^3 n \leq n$. By taking $\gamma = (\log n)^{-1/2}$, we have
\begin{equation*}
| \sup_{K,x} |t_n (K,x)| - \widetilde{Z}| = o_p (n^{-1/2 + 1/q} \max_{K} \zeta_K \log^{1+ 1/2q} n + n^{-1/6} (\max_{K} \zeta_K)^{1/3}\log^{5/6} n).
\end{equation*}

Furthermore,  $|R_1(K,x)| = o_p(a_n), |R_2(K,x)|  = o_p(a_n), |\nu_n(K,x)| = o_p(a_n)$ uniformly in $(K, x) \in \mathcal{K}_{n} \times \mathcal{X}$ with $a_n = 1/(\log n)^{1/2}$ by Lemma \ref{lemma:bounds-uniform} and the rate conditions. Again, consider the class of functions $\mathcal{F}_n  = \{ f_{n,K,x} : (K,x) \in \mathcal{K}_n \times \mathcal{X} \}$ and then
\[
E\big[ \sup_{K, x} |t_n (K,x)| \big] \lesssim \sqrt{\log n}  + (\max_K \zeta_K)^{q/(q-2)} \log n/\sqrt{n} \lesssim \sqrt{\log n} 
\]
by Lemma A.13 and Assumption \ref{assump:regularity-uniform}(\rmnum{3}), and we have $\sup_{K, x} |t_n (K,x)| \lesssim_{P} \sqrt{\log n}$.  Further, $\sup_{K, x} |Z_n (K,x)| \lesssim_{P} \sqrt{\log n}$ using Dudley's inequality (Corollary 2.2.8 in van der Vaart and Wellner (1996)) and using the same arguments given in Theorem \ref{thm:asymp-cov}, we have  $\big| \sup_{K,x}  |\widehat{T}_n(K_, x)|  -\widetilde{Z}\big|  = o_p(a_n)$ with $a_n = 1/(\log n)^{1/2}$ and 
\begin{equation}
\label{equation:coupling_Kolmogorov_That_Z}
\sup_{u \in \mathbb{R}} \big| P(\sup_{(K, x) \in \mathcal{K}_{n} \times \mathcal{X}}  |\widehat{T}_n(K_, x)| \leq u) - P(\widetilde{Z} \leq u)  \big| =  o(1).
\end{equation}

Next we consider following (infeasible) bootstrap process
\[
T_n^{e} (K,x) =\frac{\sqrt{n}(\widehat{g}_{n}^{e} (K, x) -\widehat{g}_{n} (K, x) )}{V_{n} (K, x)^{1/2}}, \quad (K,x) \in \mathcal{K}_n \times \mathcal{X}
\]
where $\widehat{g}_{n}^{e}(K, x) = \tilde{P}(K, x)' \widehat{\beta}_{K}^{e}$, $\widehat{\beta}_{K}^{e}$ is defined in \eqref{def-weightedLS} with $\tilde{P}(K,x_i)$, and $e_i$ is i.i.d. standard exponential random variables independent of $X^{n} = \{ x_1, ..., x_n\}$. Then, we have
\begin{align*}
T_n^{e} (K,x) &= \frac{\sqrt{n}(\widehat{g}_{n}^{e} (K, x) -g_0(x))}{V_{n} (K, x)^{1/2}} - \frac{\sqrt{n}(\widehat{g}_{n} (K, x) -g_0(x))}{V_{n} (K, x)^{1/2}}\\
& =  t_n^{e}(K,x) + R_n^{e}(K,x) - R_n(K,x)
\end{align*}
where $t_n^{e}(K,x) = n^{-1/2} \sum_{i=1}^{n} (e_i-1)f_{n,K,x} (\varepsilon_i, x_i)$, $R_n^{e}(K,x) = R_1^{e}(K,x) + R_2^{e}(K,x)$, $R_1^{e}(K,x)$, and $R_2^{e}(K,x)$ are defined the same as in Lemma \ref{lemma:bounds} with the rescaled data $\{ (\sqrt{e_{i}} \tilde{P}(K,x_i),\sqrt{e_{i}} \varepsilon_{i} \}_{i=1}^{n}$. Note that $\widehat{\beta}_K^{e}$ is the weighted least square estimator for the original data, and we can extend the uniform linearization results in Lemma \ref{lemma:bounds-uniform}  by replacing $\zeta_K$ with $\zeta_K^{e} = \zeta_K \log^{1/2} n $ and noting that $E[e_i] = 1, E[e_i^2] = 1, \max_{1\leq i \leq n} |e_i| = o_p(\log n)$.

By applying Theorem 2.1 in Chernozhukov et al. (2016) to the weighted bootstrap process $t_n^{e}(K,x)$, there exists a random variable $\widetilde{Z}^{e} \overset{d | X^{n}}{=} \sup_{(K, x)\in \mathcal{K}_n \times \mathcal{X}} |Z_n(K,x)|$ such that, for every $\gamma  \in (0,1)$, 
\begin{equation}
\label{equation:coupling_te_Ze}
P(|\sup_{(K, x) \in \mathcal{K}_{n} \times \mathcal{X}} |t_n^{e} (K,x)|  - \widetilde{Z}^{e}| > C_3 \delta_{2n} ) \leq C_4 (\gamma + n^{-1})
\end{equation}
where $C_{3}, C_{4}$ are positive constants that depend only on $q$, 
\begin{equation*}
\delta_{2n} = \gamma^{-1/q} n^{-1/2 + 1/q} \max_{K} \zeta_K \log^{2} n + \gamma^{-1/3}n^{-1/6} (\max_{K} \zeta_K)^{1/3} \log n,
\end{equation*}
and $\overset{d | X^{n}}{=}$ denotes that the two random variables have the same conditional distribution given $X^{n}$.

Further, 
\begin{equation*}
\big|   \sup_{K,x} |\widehat{T}_{n}^{e}(K, x)|  - \sup_{K,x} |t_{n}^{e}(K, x)| \big| \leq \sup_{K,x} |\widehat{T}_{n}^{e}(K, x) - T_n^{e}(K, x)| + \sup_{K,x} |T_{n}^{e}(K, x) - t_n^{e}(K, x)| = o_p(a_n)
\end{equation*}
by using $E\big[ \sup_{K, x} |t_n^{e} (K,x)| \big] \leq \max_{1\leq i \leq n} |e_i| E\big[ \sup_{K, x} |t_n (K,x)| \big] \lesssim_{P} \log^{3/2} n$, Assumption \ref{assump:regularity-uniform}(\rmnum{4}), and $|R_n^{e}(K,x)| = o_p(a_n), |R_n(K,x)| = o_p(a_n)$ uniformly in $(K,x) \in \mathcal{K}_n \times \mathcal{X}$ under the rate conditions in Assumption \ref{assump:regularity-uniform}(\rmnum{2}) with $a_n = 1/(\log n)^{1/2}$. Then, there exists some sequence of positive constant $\delta_{3n}, \delta_{4n}$ such that $\delta_{3n} = o(1), \delta_{4n} = o(1)$, 
\begin{equation}
\label{equation:coupling_te_thate}
P( \big|  \sup_{K,x} |\widehat{T}_{n}^{e}(K, x)|  - \sup_{K,x} |t_{n}^{e}(K, x)| \big|  > a_n \delta_{3n} ) \leq \delta_{4n}.
\end{equation}
Combining \eqref{equation:coupling_te_Ze} and \eqref{equation:coupling_te_thate} gives
\begin{equation}\label{equation:coupling_Ze_thate}
P(|\sup_{(K, x) \in \mathcal{K}_{n} \times \mathcal{X}} |\widehat{T}_{n}^{e} (K,x)|  - \widetilde{Z}^{e}| > a_n \delta_{3n} + C_3 \delta_{2n} ) \leq C_4 (\gamma + n^{-1})  + \delta_{4n}
\end{equation}
By the Markov's inequality, the following is deduced from \eqref{equation:coupling_Ze_thate}, for every $\nu \in (0,1)$, 
\begin{equation}
P(|\sup_{(K, x) \in \mathcal{K}_{n} \times \mathcal{X}} |\widehat{T}_{n}^{e} (K,x)|  - \widetilde{Z}^{e}| > a_n \delta_{3n} + C_3 \delta_{2n} |X^{n} ) \leq  \nu^{-1} (C_4 (\gamma + n^{-1})  + \delta_{4n})
\end{equation}
with probability at least $1-\nu$. Similar derivation as in Theorem \ref{thm:asymp-cov} using Lemma A.14 gives
\begin{equation}\label{equation:coupling_Kolmogorov_thate_Z_tilde}
\sup_{u \in \mathbb{R}} \big| P(\sup_{(K, x) \in \mathcal{K}_{n} \times \mathcal{X}}  |\widehat{T}_n^{e}(K_, x)| \leq u | X^{n}) - P(\widetilde{Z} \leq u)  \big| = (a_n \delta_{3n} + C_3 \delta_{2n}) \sqrt{\log n} + \nu^{-1} (C_4 (\gamma + n^{-1})  + \delta_{4n})
\end{equation}
with probability at least $1-\nu$ where we use $\widetilde{Z}^{e} \overset{d | X^{n}}{=} \widetilde{Z}$ and $E[ \sup_{K, x} |Z_n (K,x)|] \lesssim \sqrt{\log n}$. By taking $\gamma = (\log n)^{-1/2}$ and $\nu= \nu_n \rightarrow 0 $  sufficiently slower than $(\log n)^{-1/2} \vee \delta_{4n}$, and using $\delta_{2n} = o(a_n)$, the rate conditions imposed in the theorem, \eqref{equation:coupling_Kolmogorov_thate_Z_tilde} is $o_p(1)$. Combining this with \eqref{equation:coupling_Kolmogorov_That_Z}, 
\begin{equation}
\label{equation:coupling_Kolmogorov_thate_Z}
\sup_{u \in \mathbb{R}} \big| P(\sup_{(K, x) \in \mathcal{K}_{n} \times \mathcal{X}}  |\widehat{T}_n^{e}(K_, x)| \leq u | X) - P(\sup_{(K, x) \in \mathcal{K}_{n} \times \mathcal{X}}  |\widehat{T}_n(K_, x)|  \leq u)  \big| =  o_p(1).
\end{equation}
Then, the coverage result \eqref{result:coverage-uniformcb} follows. The second part of the theorem, \eqref{result:coverage-pmscb}, can be similarly derived as in the proof of Theorem \ref{thm:asymp-cov} and this completes the proof.
\end{proof}

\subsubsection{Proof of Theorem \ref{thm:PLM}}
\begin{proof}
Conditional on $X = [x_1, \cdots, x_n]'$, the following decomposition holds for any sequence $K \in \mathcal{K}_n$:
\begin{align*}
 \sqrt{n} (\widehat{\theta}_n(K) - \theta_0) = \widehat{\Gamma}_{n}(K)^{-1} S_n(K),  \quad \widehat{\Gamma}_n(K) = \frac{1}{n} (W'M_K W), \quad S_n(K) = \frac{1}{\sqrt{n}} W' M_K (g+\varepsilon)
\end{align*}
\noindent where $g = [g_1, \cdots, g_n]', g_i = g_0(x_i), g_w = [g_{w1}, \cdots, g_{wn}]', g_{wi} = g_{w0}(x_i) = E[w_i | x_i]$, and $v = [v_1, \cdots, v_n]$. All remaining proofs contain conditional expectations (conditioning on $X$) and hold almost surely (a.s.). Under Assumption \ref{assump:PLM},
\begin{align*}
\widehat{\Gamma}_{n}(K) = \Gamma_n(K) + o_p(1), \quad \Gamma_n(K) = \frac{1}{n} \sum_{i=1}^{n} M_{K, ii} E[v_i^2 | x_i]
\end{align*}
by Lemma 1 of Cattaneo, Jansson,  and Newey (2018a). Moreover,
\begin{align*}
S_n(K) =   \frac{1}{\sqrt{n}} \sum_{i=1}^{n} M_{K,ii} v_i \varepsilon_{i}  - \frac{1}{\sqrt{n}} \sum_{i=1}^{n} \sum_{j=1, j<i}^{n} P_{K,ij} (v_i \varepsilon_{j} +v_j \varepsilon_{i}) + o_p(1)
\end{align*}
since $M_{K,ij} = -P_{K, ij} $ for $j<i$, $\frac{1}{\sqrt{n}} g_w' M_K g = O_p(\sqrt{n} \overline{K}^{-\gamma_g -\gamma_{g_w}}) = o_p(1)$, $\frac{1}{\sqrt{n}} ( v' M_K g + g_w' M_K \varepsilon)  = O_p(\overline{K}^{-\gamma_g}  + \overline{K}^{-\gamma_{g_w}}) = o_p(1)$ by Lemma 2 of Cattaneo, Jansson and Newey (2018a) under Assumption \ref{assump:PLM}. Then, the following holds:
\begin{equation*}
T_n(K, \theta_0) = \sqrt{n} V_n({K})^{-1/2}(\widehat{\theta}_n(K) - \theta_0) = V_n(K)^{-1/2} \Gamma_n(K)^{-1} \frac{1}{\sqrt{n}} v' M_K \varepsilon + o_p(1) \overset{d}{\longrightarrow} N(0, 1)
\end{equation*}
by Theorem 1 of Cattaneo, Jansson and Newey (2018a).

For simplicity, here we only show the joint convergence of bivariate \textit{t}-statistics, but the proof can be easily extended to the multivariate case. For any $ K_1 < K_2$ in $\mathcal{K}_n$, we show
\begin{align}
Y_n = \Xi^{-1/2} (\delta_1 T_n(K_1,\theta_0) + \delta_2 T_n (K_2, \theta_0))  \overset{d}{\longrightarrow} N(0, 1),  \quad  \forall (\delta_1, \delta_2) \in \mathbb{R}^2
\end{align}
where $\Xi = \delta_1^2 + \delta_2^2 + 2 \delta_1 \delta_2 v_{12}, v_{12} = \lim_{n \rightarrow \infty} V_n({K_1})^{-1/2} \Gamma_{n}(K_1)^{-1} \Omega_{n}(K_1, K_2) \Gamma_{n}(K_2)^{-1} V_n({K_2})^{-1/2}$. 

Define  $Y_n  = Y_{1,n} + Y_{2,n}, Y_{1,n}$ and $Y_{2,n}$ as follows
\begin{align*}
Y_{1,n} = \omega_{1, 1n} + \sum_{i=2}^{n} y_{1,i n},  y_{1, i n} =  \omega_{1,in} + \bar{y}_{1,i n },\quad Y_{2,n} = \omega_{2, 1n} + \sum_{i=2}^{n} y_{2,i n}, y_{2, i n} =  \omega_{2,in} + \bar{y}_{2,i n },
\end{align*}
where $\omega_{1,in} = b_{1,n} W_{1,in}$, $b_{1,n} = \delta_1 \Xi^{-1/2}V_n({K_1})^{-1/2} \Gamma_{n}(K_1)^{-1}$, $W_{1,in} = v_{i} M_{K_1, i i } \varepsilon_{i} /\sqrt{n} $, $\bar{y}_{1,i n} =  \sum_{j<i} (u_{1,j}  P_{K_1, i j} \varepsilon_i + u_{1,i}  P_{K_1, i j} \varepsilon_j )/\sqrt{K_1}, u_{1,i} = c_{1,n} v_i,  c_{1,n}= -\delta_1 \Xi^{-1/2} V_n({K_1})^{-1/2} \Gamma_{n}(K_1)^{-1} \sqrt{K_{1}/n}$ and $\omega_{2,in} = b_{2,n} W_{2,in}, \bar{y}_{2,i n} $ are similarly defined with $P_{K_2}, V_n({K_2}),  \Gamma_{n}(K_2)$ and $K_2$. Note that $||V_n(K_{1})^{-1}||\leq C$ and $||\Gamma_{n}(K_{1})^{-1}||\leq C$ a.s. for $n$ large enough by Assumption \ref{assump:PLM}, and it follows that $|| b_{1,n} || \leq C$. Also, $E[ ||\omega_{1,1n}||^{4} | X] \leq C\sum_{i=1}^{n} E[||W_{1,in}||^{4} | X] \rightarrow 0$ a.s. by Assumption \ref{assump:PLM}(ii). Using the same arguments in the proof of Lemma A2 in Chao et al. (2012), we have $\omega_{1,1n} = o_p(1)$ and $\omega_{2,1n} = o_p(1)$ unconditionally,  thus $Y_n =\sum_{i=2}^{n} y_{in} + o_p(1), y_{in} = y_{1, i n} +  y_{2, i n}$. 

Let $\mathcal{X}_{i} = (W_{1,in}, W_{2,in}, v_i, \varepsilon_{i})^{\prime}$ and define the $\sigma$-fields $F_{i,n}= \sigma(\mathcal{X}_{1}, ..., \mathcal{X}_{i})$ for $i=1, ..., n.$ Then, conditional on $X$, $\{y_{in}, F_{i,n}, 1\leq i \leq  n, n \geq 2\}$ is a martingale difference array with $F_{i-1,n} \subseteq F_{i, n}$. We apply the martingale central limit theorem to show, conditional on $X$, $\sum_{i=2}^{n} y_{in}  \overset{d}{\longrightarrow} N(0, 1)$ a.s.  Note that $E[\omega_{1,in} \bar{y}_{1,jn}|X] = 0, E[\omega_{1,in} \bar{y}_{2,jn}|X] = 0, E[\omega_{2,in} \bar{y}_{1,jn}|X] = 0, E[\omega_{2,in} \bar{y}_{2,jn}|X] = 0$ for all $i,j$. Then similar  to the proof of Lemma A2 in Chao et al. (2012),
\begin{align*}
 s_{n}^{2}(X) &= E [(\sum_{i=2}^{n} y_{in})^2 |X ] = \sum_{i=2}^{n} ( E[\omega_{1,in}^2 | X] + E[\bar{y}_{1,in}^2 | X] ) + \sum_{i=2}^{n} ( E[\omega_{2,in}^2 | X] + E[\bar{y}_{2,in}^2 | X] ) \\
&+ 2 \sum_{i=2}^{n} ( E[\omega_{1,in}\omega_{2,in} | X] + E[\bar{y}_{1,in}\bar{y}_{2,in} | X])\\
&= \delta_{1}^{2}\Xi^{-1} + \delta_{2}^{2}\Xi^{-1}  - E[\omega_{1,1n}^2 | X] -  E[\omega_{2,1n}^2 | X] - 2 E[\omega_{1,1n} \omega_{2,1n}| X] \\
& + 2 \delta_1 \delta_2 \Xi^{-1} V_n({K_1})^{-1/2} \Gamma_{n}(K_1)^{-1} \Omega_{n}(K_1, K_2) \Gamma_{n}(K_2)^{-1} V_n({K_2})^{-1/2} \rightarrow 1 \quad a.s.
\end{align*}
Moreover, we have $\sum_{i=2}^n E[y_{in}^4|X] \lesssim  \sum_{i=2}^n E[y_{1, i n}^{4}|X] + \sum_{i=2}^n E[y_{2, i n}^{4}|X] \overset{a.s.}{\rightarrow} 0$ as in the proof of Lemma A2 of Chao et al. (2012).

 It remains to prove that for any $\delta>0$, $P(\big| \sum_{i=2}^{n} E[y_{in}^2| \mathcal{X}_1,..., \mathcal{X}_{i-1}, X] - s_{n}^{2}(X) \big| \geq \delta  | X) \rightarrow 0$. Note that
\begin{align}
&\nonumber\sum_{i=2}^{n} E[y_{in}^{2}| \mathcal{X}_1,..., \mathcal{X}_{i-1}, X] - s_{n}^{2}(X)  \\
&= \sum_{i=2}^{n} E[y_{1, in}^{2}| \mathcal{X}_1,..., \mathcal{X}_{i-1}, X] -  \sum_{i=2}^{n} ( E[\omega_{1,in}^2 | X] + E[\bar{y}_{1,in}^2 | X] ) \label{equation:plmproof_mclt1}\\
&+ \sum_{i=2}^{n} E[y_{2, in}^{2}| \mathcal{X}_1,..., \mathcal{X}_{i-1}, X]-  \sum_{i=2}^{n} ( E[\omega_{2,in}^2 | X] + E[\bar{y}_{2,in}^2 | X] )\label{equation:plmproof_mclt2}\\
&+ 2 \Big( \sum_{i=2}^{n} (E[\omega_{1, in}\omega_{2, in}| \mathcal{X}_1,..., \mathcal{X}_{i-1}, X] - E[\omega_{1,in}\omega_{2,in} | X])  \label{equation:plmproof_mclt3}\\
&+ \sum_{i=2}^{n} E[\omega_{1,in} \bar{y}_{2,in} + \omega_{2,in} \bar{y}_{1,in} |\mathcal{X}_1,..., \mathcal{X}_{i-1}, X ]+  \sum_{i=2}^{n} (E[\bar{y}_{1,in}\bar{y}_{2,in} | \mathcal{X}_1,..., \mathcal{X}_{i-1}, X] - E[\bar{y}_{1,in}\bar{y}_{2,in} | X] ) \Big)\label{equation:plmproof_mclt4}.
\end{align}
\eqref{equation:plmproof_mclt1} and \eqref{equation:plmproof_mclt2} converge to 0 a.s. by the proof of Lemma A2 in Chao et al. (2012). Moreover, it is straightforward to verify that \eqref{equation:plmproof_mclt3} and \eqref{equation:plmproof_mclt4} converge to 0 a.s. since $P_{K_1, i j} P_{K_2, i j} \leq P_{K_1, i j}^{2}  \vee P_{K_2, i j}^{2}$, $K_{1} \asymp  K_{2}$ and  by closely following the proof of Lemma A2 in Chao et al. (2012). Then we can apply the martingale central limit theorem and deduce $Y_n \overset{d}{\longrightarrow} N(0, 1)$ using similar arguments to the proof of Lemma A2 in Chao et al. (2012).  Coverage results \eqref{result:coverage-plmci} and \eqref{result:coverage-plmci-pms} follow by the joint convergence of $\widehat{T}_{n}(K, \theta_0)$ with $\max \limits_{K \in \mathcal{K}_n} |\frac{\widehat{V}_n (K)}{V_n (K)} -1|  = o_p(1), || \widehat{\Sigma}_n  - \Sigma_n||  = o_p(1)$ as $n, K \rightarrow \infty$ under the assumption imposed in Theorem \ref{thm:PLM} and the Slutzky theorem. This completes the proof. 
\end{proof}

\newpage

\section{Figures and Tables}
\label{sec:figtable}

\newpage

\renewcommand{\tabcolsep}{4pt}
\begin{table}[htbp]
\centering
\begin{threeparttable}
   \caption{Coverage and Length of Nominal 95$\%$ CIs and CBs - Splines}
    \label{table:cov-spline-n1000}
    \begin{tabular}{l ccc ccc ccc ccc cc}
    \toprule
    & \multicolumn{11}{c}{Pointwise} & & \multicolumn{2}{c}{Uniform}\\
    \cmidrule{2-12} \cmidrule {14-15}
     & \multicolumn{2}{c}{$x=0.2$}  &  &  \multicolumn{2}{c}{$x=0.5$} & & \multicolumn{2}{c}{$x=0.8$}  & & \multicolumn{2}{c}{$x=0.9$} & &  \multicolumn{2}{c}{}\\
      \cmidrule{2-3}  \cmidrule{5-6} \cmidrule{8-9} \cmidrule{11-12}
          & COV & AL  && COV & AL && COV & AL && COV & AL &&   COV & AL\\
 \midrule      
    \multicolumn{9}{l}{Model 1: $g_1(x) = \ln (|6x-3| + 1)sgn(x-1/2)$}\\ 
  \ \ \ Standard   & 0.93 & 0.27 && 0.93 & 0.36 && 0.91 & 0.92 && 0.92 & 1.49  && 0.42 & 0.69\\
    \ \ \ Robust ($\widehat{K}_{\texttt{cv}}$)  & 0.98 & 0.37 && 0.98 & 0.46 && 0.96 & 1.14 && 0.95 & 1.76  && 0.97 & 1.33\\
    \ \ \ Robust ($\widehat{K}_{\texttt{cv+}}$) & 0.98 & 0.51 && 0.98 & 0.49 && 0.98 & 1.51 && 0.97 & 2.08  && 0.98 & 1.42\\
  \\
       \multicolumn{9}{l}{Model 2: $g_2(x) =  \sin (7\pi x/2)/ [1+ 2x^2(sgn(x)+1)]$}\\     
  \ \ \ Standard   & 0.80 & 0.28 && 0.93 & 0.36 && 0.91 & 0.92 && 0.92 & 1.49  && 0.27 & 0.69\\
    \ \ \ Robust ($\widehat{K}_{\texttt{cv}}$)  & 0.93 & 0.37 && 0.97 & 0.46 && 0.96 & 1.14 && 0.95 & 1.76  && 0.96 & 1.33\\
    \ \ \ Robust ($\widehat{K}_{\texttt{cv+}}$) & 0.98 & 0.51 && 0.98 & 0.49 && 0.98 & 1.51 && 0.97 & 2.08  && 0.98 & 1.42\\
  \\
 \multicolumn{9}{l}{Model 3: $g_3(x) = x-1/2 + 5 \phi(10(x-1/2))$}\\     
  \ \ \ Standard   & 0.77 & 0.29 && 0.65 & 0.40 && 0.89 & 1.00 && 0.91 & 1.57  && 0.16 & 0.70\\
    \ \ \ Robust ($\widehat{K}_{\texttt{cv}}$)  & 0.88 & 0.39 && 0.74 & 0.50 && 0.96 & 1.23 && 0.95 & 1.85  && 0.75 & 1.35\\
    \ \ \ Robust ($\widehat{K}_{\texttt{cv+}}$) & 0.98 & 0.52 && 0.92 & 0.53 && 0.98 & 1.52 && 0.97 & 2.06  && 0.97 & 1.44\\
  \\
          \bottomrule
    \end{tabular}
     \begin{tablenotes}
    \small
    \item[] Notes: ``Pointwise'' reports coverage (COV) and average length (AL) of (1) the standard 95\% CI with $\widehat{K}_{\texttt{cv}} \in \mathcal{K}_n$; (2) robust CI with $\widehat{K}_{\texttt{cv}}$; (3) robust CI with $\widehat{K}_{\texttt{cv+}}$. ``Uniform" reports analogous uniform inference results for confidence bands.  $\widehat{K}_{\texttt{cv}}$ is selected to minimize leave-one-out cross-validation and $\widehat{K}_{\texttt{cv+}} = \widehat{K}_{\texttt{cv}} +2$. Using quadratic spline regressions with evenly placed knots.
    \item[] 
    \end{tablenotes}
  \end{threeparttable}
\end{table}%

\newpage

\renewcommand{\tabcolsep}{6pt}

\begin{table}[htbp]
\centering
\begin{threeparttable}
   \caption{Nonparametric Wage Elasticity of Hours of Work Estimates in Blomquist and Newey (Table 1, 2002). Wage elasticity evaluated at the mean net wage rates, virtual income, and level of hours.}
    \label{table:BN}
    \begin{tabular}{cccccc}
    \toprule
    &Additional Terms\tnote{1} & $CV\tnote{2}$ & $\widehat{E}_{w}$ & $SE_{\widehat{E}_{w}}$ & $CI_{\widehat{E}_{w}} (K)$ \\
      \midrule
   &$1, y_J, w_J$ & 0.00472 &    0.0372  & 0.0104 &  [0.0168, 0.0576]  \\
    &$\Delta y \Delta w$& 0.0313 &   0.0761 &   0.0128 & [0.0510, 0.1012] \\
  & $\ell \Delta y$& 0.0305 &  0.0760   & 0.0127 &  [0.0511, 0.1009]\\
   & $y_J^2, w_J^2$ & 0.0323 &  0.0763   & 0.0129 & [0.0510, 0.1016] \\
   & $\Delta y^2, \Delta w^2 $ &  0.0369&    0.0543  &  0.0151  &  [0.0247, 0.0839]\\
   & $y_J w_J$ & 0.0364 &  0.0659 &   0.0197   & [0.0273, 0.1045] \\
   &$\Delta yw $ & 0.0350 &   0.0628   & 0.0223  & [0.0191, 0.1065] \\
    &$\ell^2 \Delta y$ & 0.0364 &   0.0636  &  0.0223 &[0.0199, 0.1073]  \\
   & $y_J^3, w_J^3$ & 0.0331 &   0.0845  &  0.0275&   [0.0306, 0.1384] \\
  & $\ell \Delta y^2, \ell \Delta w^2, \ell \Delta y w$ & 0.0263 & 0.0775 &   0.0286  & [0.0214,     0.1336] \\
   & $y_J^2w_J, y_J w_J^2$ & 0.0252 &     0.0714   & 0.0289& [0.0148, 0.1280]\\
    \midrule
   & MLE estimates &  &    0.123   & 0.0137 & \\
    \midrule
        \multicolumn{6}{l}{ critical values: $\widehat{c}_{1-\alpha}(x) =2.503$, \quad $CI_{\widehat{E}_w}^{\texttt{sup}} (\widehat{K}_{\texttt{cv}})=[0.0165, 0.0921]$\tnote{3} } \\
      \multicolumn{6}{l}{$CI_{\widehat{E}_w}^{\texttt{sup}} (\widehat{K}_{\texttt{cv+}})= [0.0166, 0.1152], \quad CI_{\widehat{E}_w}^{\texttt{sup}} (\widehat{K}_{\texttt{cv++}})  = [0.0070, 0.1186]$} \\
          \bottomrule
    \end{tabular}
    \vspace{0.3cm}
    \begin{tablenotes}
    \small
    \item[1] $y$ : non-labor income,  $w$ : marginal wage rates, $\ell$: the end point of the segment in a piecewise linear budget set. $\ell^m \Delta y^p w^q$ denotes $\sum_j \ell_j^{m} (y_j^{p} w_j^{q} - y_{j+1}^{p} w_{j+1}^{q})$.
    \item[]
\item[2] $CV$ denotes the cross-validation criteria defined in Blomquist and Newey (2002, p.2464). $\widehat{K}_{\texttt{cv}} = K_5$, the 5th smallest model, is chosen by the cross-validation, and let $\widehat{K}_{\texttt{cv+}} = K_6$, $\widehat{K}_{\texttt{cv++}} = K_7$. 
\item[3] $CI_{\widehat{E}_w}^{\texttt{sup}} (K) = \widehat{E}_w (K) \pm \widehat{c}_{1-\alpha}(x) SE_{\widehat{E}_w} (K)$, $CI_{\widehat{E}_{w}} (K) = \widehat{E}_w (K) \pm z_{1-\alpha/2} SE_{\widehat{E}_w} (K)$.
    \end{tablenotes}
  \end{threeparttable}
\end{table}%

\newpage

\newpage

\end{appendices}

\end{document}